\documentclass[journal]{IEEEtran}
\ifCLASSINFOpdf
\usepackage[pdftex]{graphicx}
\else
\fi
\usepackage[cmex10]{amsmath}
\usepackage[tight,footnotesize]{subfigure}
	\usepackage{amsfonts}
	\usepackage{graphics} 
	\usepackage{setspace}
	\usepackage{amssymb}
	\usepackage{epsfig} 
	\usepackage{stfloats}
	\usepackage{multirow}
	\usepackage{xcolor}
	\usepackage{makecell}
	\usepackage{amsmath}
	\usepackage{booktabs}  
	\usepackage{array}     
	\usepackage{algorithm}
	\usepackage{algorithmicx}
	\usepackage{algpseudocode}
	\usepackage{graphicx} 
	\usepackage{amsthm}
	\floatname{algorithm}{Algorithm}
	\usepackage{enumerate}
	\usepackage{subeqnarray}
	\theoremstyle{plain}

	\usepackage{bm}
	
\begin{document}
		%
		\title{Graph-Aware Temporal Encoder Based \\Service Migration and Resource Allocation\\ in Satellite Networks}

		\author{Haotong~Wang,~\IEEEmembership{Graduate Student Member,~IEEE}, Jun~Du,~\IEEEmembership{Senior Member,~IEEE}, Chunxiao~Jiang,~\IEEEmembership{Fellow,~IEEE}, \\Jintao~Wang,~\IEEEmembership{Senior Member,~IEEE}, M\'{e}rouane~Debbah,~\IEEEmembership{Fellow,~IEEE}, and Zhu~Han,~\IEEEmembership{Fellow,~IEEE}

			\thanks{Haotong Wang is with the Department of Electronic Engineering, Tsinghua University, Beijing 100084, China. (e-mail: wht24@mails.tsinghua.edu.cn).}
		\thanks{Jun Du and Jintao Wang are with the Department of Electronic Engineering, Tsinghua University, Beijing 100084, China, and also with the State Key Laboratory of Space Network and Communications, Tsinghua University, Beijing 100084, China. (e-mail: jundu@tsinghua.edu.cn, wangjintao@tsinghua.edu.cn).}
		\thanks{Chunxiao Jiang is with Beijing National Research Center for Information Science
			and Technology, Tsinghua University, Beijing 100084, China, and also with the State Key Laboratory of Space Network and Communications, Tsinghua University, Beijing 100084, China. (e-mail: jchx@tsinghua.edu.cn).}
		\thanks{M\'{e}rouane Debbah is with KU 6G Research Center, Khalifa University of Science and Technology, PO Box 127788, Abu Dhabi, UAE. (email: merouane.debbah@ku.ac.ae).}
		\thanks{Zhu Han is with the Department of Electrical and Computer Engineering at the University of Houston, Houston, TX 77004 USA, and also with the Department of Computer Science and Engineering, Kyung Hee University, Seoul, South Korea, 446-701. (e-mail: hanzhu22@gmail.com).}

		}

		
		%


		\maketitle
		
		\begin{abstract}
		The rapid expansion of latency-sensitive applications has sparked renewed interest in deploying edge computing capabilities aboard satellite constellations, aiming to achieve truly global and seamless service coverage. On one hand, it is essential to allocate the limited onboard computational and communication resources efficiently to serve geographically distributed users. On the other hand, the dynamic nature of satellite orbits necessitates effective service migration strategies to maintain service continuity and quality as the coverage areas of satellites evolve.
		We formulate this problem as a spatio-temporal Markov decision process, where satellites, ground users, and flight users are modeled as nodes in a time-varying graph. The node features incorporate queuing dynamics to characterize packet loss probabilities.
		To solve this problem, we propose a Graph-Aware Temporal Encoder (GATE) that jointly models spatial correlations and temporal dynamics. GATE uses a two-layer graph convolutional network to extract inter-satellite and user dependencies and a temporal convolutional network to capture their short-term evolution, producing unified spatio-temporal representations. The resulting spatial-temporal representations are passed into a Hybrid Proximal Policy Optimization (HPPO) framework. This framework features a multi-head actor that outputs both discrete service migration decisions and continuous resource allocation ratios, along with a critic for value estimation.
		We conduct extensive simulations involving both persistent and intermittent users distributed across real-world population centers. 
		The results validate that the proposed framework consistently achieves superior performance compared to Proximal Policy Optimization (PPO), Soft Actor Critic (SAC), and ablated baselines in terms of reward, failure rate, and migration overhead, demonstrating the effectiveness of the proposed spatio-temporal modeling and hybrid reinforcement learning approach in dynamic satellite edge environments.
		\end{abstract}
		\vspace{-1mm}

		%
		\IEEEpeerreviewmaketitle

		\vspace{-2mm}
		\section{Introduction}

		The explosive growth of smart devices and latency‐sensitive applications, ranging from real‐time video streaming to autonomous vehicle control, has placed unprecedented demands on modern communications infrastructures. While terrestrial cellular networks provide high capacity in urban centers, they struggle to deliver seamless, low‐latency service in rural, maritime, and other remote regions~\cite{fontanesi2025artificial}. Traditional cloud computing architectures compound this challenge by introducing long round‐trip times between geographically dispersed users and distant data centers~\cite{tang2024joint}. 
		In response, the satellite edge computing paradigm has emerged as a promising solution for achieving truly global service coverage \cite{9444334}. By equipping satellites with onboard computing and storage resources, it is possible to integrate the data plane and control plane within a unified platform. 
		Unlike conventional ``bent-pipe'' satellites that merely relay traffic to ground stations, edge-enabled satellites can process user tasks directly in orbit. This significantly reduces end-to-end latency and alleviates congestion in the backhaul network. Major industry initiatives such as Starlink, OneWeb, and Guowang are already deploying large constellations of Low Earth Orbit (LEO) and Medium Earth Orbit (MEO) satellites, highlighting the strategic value of in-orbit edge services~\cite{wang2024resource}.

		Despite these advances, orchestrating service placement aboard fast‐moving satellites poses fundamental \textbf{challenges}~\cite{gao2024hierarchical}. 
		First, the mobility of satellites leads to continuous changes in network topology~\cite{mao2024intelligent}. Each satellite dynamically serves different subsets of users as it moves along its orbit, requiring active service instances to migrate in order to maintain user connectivity. 
		Second, onboard resources such as compute cycles, memory, and downlink bandwidth are severely constrained due to limitations in power, weight, and thermal capacity~\cite{zhao2025demand}. Third, user service demand is heterogeneous~\cite{maity2024traffic}. While stationary Ground Users (GUs) generate persistent task streams, Flight Users (FUs) (e.g., commercial aircraft) produce bursty requests only during stable cruising phases, leading to highly variable traffic patterns.
		
		Given the highly dynamic topology and heterogeneous user demands described above, ensuring uninterrupted connectivity requires service migration, defined as the process of transferring an active service session from one satellite to another. However, it introduces nontrivial overhead~\cite{zakarya2024apmove}. Migration entails state transfer overhead over limited inter‐satellite or satellite‐ground links, potential service interruption during handoff, and additional compute and energy consumption. Excessive migrations degrade user experience and exhaust precious in‐orbit resources, while insufficient migrations risk coverage gaps and service failures~\cite{ji2024dynamic}.
		
		To manage these trade-offs effectively, a decision-making mechanism must jointly optimize satellite selection, resource allocation, and migration timing under dynamic spatial and temporal conditions. This mechanism needs to account for the satellite coverage changes over time, the varying computational and communication resources onboard, and the bursty and heterogeneous service requests from different users. Hence, we propose a Graph-Aware Temporal Encoder (GATE) combined with Hybrid Proximal Policy Optimization (HPPO) for service migration and resource allocation.

		\vspace{-1mm}
		\subsection{Related Work and Motivation}
		 Early research on satellite networks primarily focused on transparent payload architectures, where data is relayed to terrestrial data centers without onboard processing~\cite{9444334,du2022sdnTON}. More recent studies emphasize a shift toward equipping satellites with onboard computing and edge services, enabling reduced latency and enhanced scalability.
		
		Concurrently, Federated Learning (FL) frameworks have been adapted to the satellite environment by leveraging inter-satellite links (ISLs) to address intermittent connectivity.
		In \cite{jiang2025federated}, a FL framework is proposed for mobile traffic prediction in satellite-terrestrial networks, leveraging an adaptive graph convolutional network and LSTM to achieve a balance between data privacy and prediction accuracy. 
		The author of \cite{razmi2024board} exploits the predictable nature of satellite orbits to enable synchronous or asynchronous parameter aggregation, achieving a seven-fold improvement in convergence speed and a 90\% reduction in communications overhead. 
		To resolve hardware heterogeneity and model staleness, \cite{lin2024fedsn} introduces dynamic sub-model training and pseudo-synchronous aggregation, leading to improved accuracy with reduced overhead.
		Further advancing decentralization, \cite{yang2024dfedsat} incorporates dual adaptive mechanisms for intra- and inter-plane model propagation, along with self-compensation for ISL failures, demonstrating robust convergence under dynamic topologies. 
		For mega-constellations, \cite{shi2024satellite} integrates ring-based intra-orbit aggregation with network-flow-based global transmission to reduce reliance on ground stations and accelerate convergence by 30\% . 
		In \cite{gong2024multi}, the authors propose a Lyapunov-based multi-modal FL algorithm and a blockchain verification protocol, which jointly optimize resource allocation and enhance security.
		Collectively, these efforts highlight the importance of cross-satellite cooperation and communication-efficient FL in enabling intelligence for future satellite networks.
		
		Recent advances in satellite-edge computing have addressed diverse challenges in distributed task processing.
		\cite{zhang2024collaborative} proposes a multi-agent task offloading scheme for distributed satellite mobile edge computing, leveraging counterfactual multi-agent policy gradients with attention-based bidirectional long short-term memory networks (BiLSTM) under a Centralized Training with Decentralized Execution (CTDE) framework to optimize energy-efficient decisions amid spatio-temporal constraints. 
		\cite{zhao2024qos} proposes a Quality of Service (QoS)-aware multihop offloading algorithm for satellite-terrestrial networks, integrating genetic algorithms with Lagrangian methods to minimize computation latency while balancing workloads across visible and distant satellites. 
		\cite{zhou2024latency} proposes a dual-cloud edge collaboration architecture coordinated by Geostationary Earth Orbit (GEO) satellites and ground clouds, utilizing Generalized Proximal Policy Optimization (GePPO)-based deep Reinforcement Learning (RL) to adaptively minimize latency and energy consumption in dynamic LEO environments. 
		\cite{zhong2025joint} proposes a customized task utility model that incorporates heterogeneous user demand. They formulate joint task offloading and resource allocation as a mixed-integer nonlinear programming problem, solved using a multilayer optimization framework that integrates successive convex approximation and RL.
		In~\cite{gong2025multi}, a multi-task integrated computation offloading model is proposed for satellite-ground integrated networks, which enhances data transmission rates, privacy levels, and resource optimization under dynamic network conditions. 
		The work in~\cite{cai2024dynamic} proposes a Graphic Deep Reinforcement Learning (GDRL) approach that integrates graph neural networks for spatial feature extraction and meta-learning for fast adaptation.
		The authors of~\cite{du2025collaborative} develop a graph-based collaborative offloading strategy for LEO satellite networks that dynamically allocates tasks by analyzing link and on-board resources, significantly reducing computing latency.

		The mobility of satellites inevitably disrupts service continuity, making adaptive service migration essential for maintaining consistent task execution across dynamically changing satellite visibility~\cite{duj2025CM}.
		The authors of~\cite{wang2024dynamic} propose a privacy-preserving migration framework for Satellite Edge Computing that combines latency, migration cost, and location privacy leakage into a multi-objective cost function, solved via improved Monte-Carlo control to prevent user tracking. 
		\cite{chen2024spaceedge} introduces a deep reinforcement learning approach that jointly optimizes service migration and power control in energy-harvesting LEO satellites, achieving superior latency performance through both centralized and multi-agent learning. 
		For mega-constellations, \cite{li2023online} leverages convolutional proximal policy optimization to dynamically deploy services while balancing migration costs and delay satisfaction ratios. 
		To overcome the state space explosion in migration decisions, \cite{li2021distributed} develops a distributed two-layer decomposition model that reduces dimensionality, significantly improving decision efficiency. 
		In addition, bandwidth-aware migration combines online lazy migration with virtual CPU scheduling to achieve competitive performance guarantees while meeting task deadlines through hierarchical resource decoupling \cite{deng2023bandwidth}. 
		
		Despite recent progress, most studies address isolated aspects of satellite-edge orchestration: some focus on static placement or oversimplified mobility models, others optimize allocation or migration in isolation. In real networks these aspects are tightly coupled, GUs require persistent, reliable connectivity while FUs are intermittent and highly mobile, so migrations must be invoked judiciously to preserve coverage without overloading scarce onboard resources. Table \ref{tab:related} summarizes representative works across objective, system model, spatio-temporal modeling, and action space, showing prior art typically treats prediction, offloading, or energy-aware migration separately. By contrast, our work jointly models queue dynamics and packet loss on a time-varying graph, uses GATE for spatio-temporal encoding, and learns a hybrid policy enabling queue-aware migration decisions that balance service continuity and migration cost in dynamic satellite networks.

\begin{table*}[!t]
	\small
	\centering
	\caption{Comparison of representative related works in satellite edge computing}
	\vspace{-2mm}
	\label{tab:related}
	\begin{tabular}{p{0.8cm} p{4.8cm} p{3.8cm} p{2.8cm} p{3.2cm}}
		\toprule
		\textbf{Work} & \textbf{Optimization Objective(s)} & \textbf{System Model} & \textbf{Spatio-temporal Modeling} & \textbf{Action space} \\
		\midrule
		\cite{jiang2025federated} & Privacy-aware traffic prediction & Satellite–ground networks & GCN \& LSTM & -(supervised prediction) \\
		\midrule
		\cite{lin2024fedsn} & Accuracy with reduced overhead & FL with LEO constellation & Pseudo-synchronous aggregation & -(training/aggregation) \\
		\midrule
		\cite{cai2024dynamic} & Maximize resource utilization & Dynamic SAGIN and dual service types& GNN for feature extraction & Sequential task offloading \& resource allocation \\
		\midrule
		\cite{du2025collaborative} & Minimize task computing latency & Satellites with time-varying topology & Computing Coordinate Graph & Node selection \& task partitioning \\
		\midrule
		\cite{chen2024spaceedge} & Minimize latency under energy constraints & Energy-harvesting LEO & MDP with temporal dynamics & Service migration (discrete) \& power control (continuous) \\
		\midrule
		\textbf{This work} & \textbf{Maintain service continuity \& QoS in time-varying constellations} & \textbf{SAGIN with heterogeneous users and queue dynamics } & \textbf{Graph-Aware Temporal Encoder} & \textbf{Discrete service migration \& continuous resource allocation ratios} \\
		\bottomrule
	\end{tabular}
	\vspace{-4mm}
\end{table*}
		\vspace{-6mm}
		\subsection{Contributions}
		\vspace{-1mm}
		In this paper, we propose GATE-HPPO, a framework that models the satellite–user system as a time-varying graph and learns joint policies for service migration and resource allocation.
		Unlike prior works that employ Graph Neural Networks (GNNs), temporal models, or RL independently, our framework establishes an integrated spatio-temporal decision mechanism tailored to dynamic satellite networks. 
		It enables coordinated learning of both spatial correlations and temporal queue evolution, allowing the policy to adaptively balance service continuity and migration overhead under time-varying connectivity. 
		The detailed algorithmic design of GATE-HPPO is presented in the following sections, and our main contributions are summarized as follows:
		\begin{enumerate}
			\item We propose a unified optimization framework for satellite networks that jointly integrates resource allocation and service migration. To capture user heterogeneity, we model two distinct traffic types. GUs remain stationary, generating persistent, high-volume, and delay-sensitive demands throughout the entire time horizon. In contrast, FUs follow predefined trajectories and initiate requests only during stable cruising phases, typically with lighter and more sporadic workloads. A queuing-based formulation is employed to represent these diverse service patterns while accounting for onboard constraints in CPU, bandwidth, and migration overhead, providing a practical foundation for informed decision-making.

			\item To tackle the dynamic and structured nature of the problem, we embed the satellite–user environment into a sequence of time‐varying graphs. These graphs, along with their associated queueing states, are transformed into Markov Decision Process (MDP) states. This formulation enables the application of a deep reinforcement learning approach to jointly optimize discrete satellite selection and continuous resource allocation. The spatio-temporal MDP abstraction allows the learned policy to adapt effectively to both changes in spatial connectivity and fluctuations in user demand over time.

			\item We develop GATE‐HPPO, a unified learning architecture that tightly couples spatio-temporal representation and hybrid decision optimization into a cohesive, end-to-end learning pipeline. 
			Specifically, GATE employs Graph Convolutional Networks (GCNs) to model dynamic inter-satellite and user connectivity, while Temporal Convolutional Networks (TCNs) capture queue evolution and migration trends over time. 
			The fused embeddings yield a coherent state representation that is directly optimized by HPPO, which unifies discrete migration control and continuous resource allocation within a single optimization process. 
			Through this end-to-end coupling of representation and decision learning, GATE‐HPPO transcends modular combinations and achieves stable, low-overhead service migration in dynamic satellite constellations.

			\item  Through extensive simulations under realistic satellite constellations and heterogeneous user distributions, including both ground and aerial clients, we demonstrate that GATE‐HPPO consistently outperforms baseline methods across key metrics. Our method achieves higher accumulated reward, lower service failure rates, and fewer service migrations, validating its effectiveness for operational satellite edge computing.
		\end{enumerate}

		\vspace{-3mm}
		\subsection{Organization}
		The remainder of this paper is structured as follows. 
		Section~\ref{sys} describes the system model. Section~\ref{pro_for} introduces the problem formulation.
		Section~\ref{scheme} details the GATE-HPPO algorithm. Section~\ref{sim} presents the simulation results and discussion. Finally, conclusions are drawn in Section~\ref{conculsion}.

		\vspace{-1mm}
		\section{System Model}
		\vspace{-1mm}
		\label{sys}
		\begin{figure}[t]
			\centering
			\includegraphics[width=0.45\textwidth]{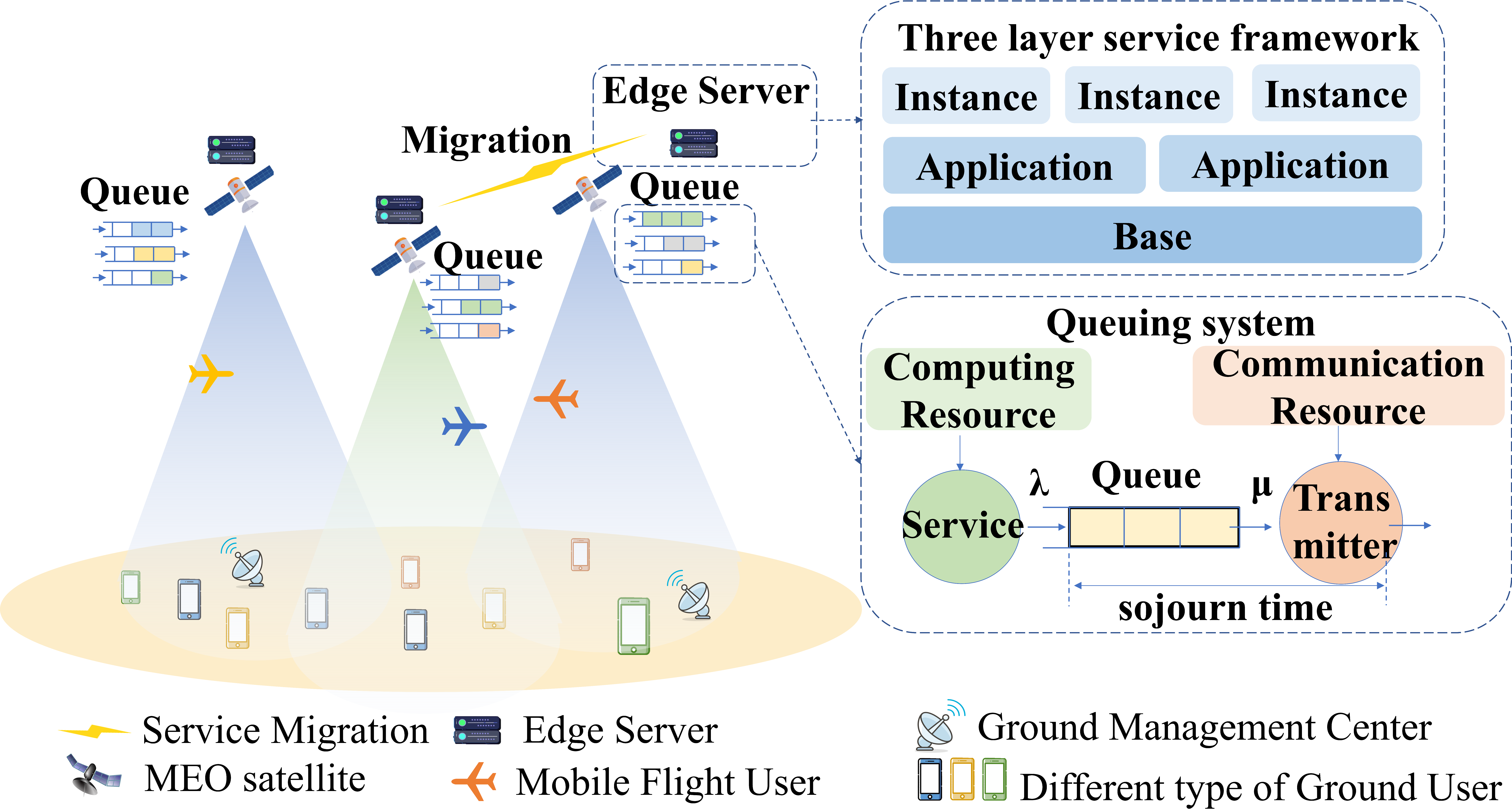}\vspace{-2mm}
			\caption{Seamless satellite service scenarios.}\vspace{-4mm}
			\label{fig:1}
		\end{figure}
	    \subsection{Network Model}
		We consider a satellite network, which consists of $S$ satellites denoted as $\mathcal{S}=\{1,2,3,...,S\}$.
		We let $\mathcal{U}$ denote the complete set of users, which is partitioned into static GUs and mobile FUs: $\mathcal{U}=\mathcal{U}_1 \cup \mathcal{U}_2$, where $\mathcal{U}_1=\{1,2,3,...,U_1\}$, represents $U_1$ GUs with fixed locations, and $\mathcal{U}_2=\{1,2,3,...,U_2\}$, represents $U_2$ FUs whose service demand arises only during level, non‑maneuvering flight.
		As shown in Fig.~\ref{fig:1}, the satellites are deployed as edge servers, providing seamless services for GUs and FUs during a specific period $T$. To support users' demands, each satellite has to allocate computing resources to run applications and allocate communication resources to transmit. As the mobility of FUs and satellites, it is necessary to migrate applications during the handover process for continuous service.
		There are $C$ Ground Management Centers (GMCs) interconnected via fiber links and can be logically regarded as a unified management center. It is responsible for tracking, telemetry, and command operations, providing essential satellite management and control to maintain normal satellite operations in orbit. Based on ephemeris data and flight schedules, GMCs can observe the global information and make resource allocation and service migration decisions.
		We assume that the total period $T$ can be divided into $N$ time slots, each slot length $\delta$ is $T/N$, which is sufficiently small so that the satellite positions and channel conditions remain constant within each time slot. The main notations used throughout the following sections are summarized in Table~\ref{tab:parameters} for convenience. 
		
		\begin{table}[!t]
			\small
			\caption{LIST OF MAIN NOTATIONS\label{tab:table1}}
			\vspace{-3mm}
			\centering
			\begin{tabular}{c|p{6.5cm}}
				\hline
				\hline
				\textbf{Parameter} & \textbf{Definition} \\
				\hline
				$S$       & Number of satellites \\
				$U_1$       & Number of ground users \\
				$U_2$       & Number of mobile flight users \\
				$N$       & Number of time slots \\
				$\delta$  & Length of per time slot (s) \\
				$\gamma_{u,s}(n)$ & SNR between user $u$ and satellite $s$ at time $n$ \\
				$P_{TX}$  & Transmission power of satellite (W)\\
				$G_{TX}$  &Transmitter antenna gain of satellite \\
				$L_{TX}$  &Transmitter feeder loss of satellite \\
				$G_{RX}$  &Receiver antenna gain of user \\
				$F_{RX}$  &Receiver equivalent noise temperature (K)\\
				$\Theta$  &Boltzmann constant (J/K)\\
				$W_{RF}$  &Reference bandwidth (Hz)\\
				$L_{u,s}(n)$ &Propagation loss\\
				$L^f_{u,s}(n)$ &Free-space path loss\\
				$L^r_{u,s}(n)$ &Rain attenuation\\
				$\theta_{u,s}(n)$ &Elevation angles ($^{\circ}$)\\
				$\theta_{min}$ &Minimum elevation angle ($^{\circ}$)\\
				\hline
				\hline
			\end{tabular}
			\label{tab:parameters}
			\vspace{-3mm}
		\end{table}
		\vspace{-2mm}		
		\subsection{Communication Model}
		\subsubsection{Channel Model}
		In our system, to ensure continuous, high‐reliability links between user terminals and satellites, each user is equipped with a single antenna while each satellite employs a multi‐element array to facilitate beamforming and spatial multiplexing~\cite{cao2022edge}. 
		To provide orthogonal channel access, we adopt an Orthogonal Frequency-Division Multiple Access (OFDMA) scheme, in which the total bandwidth is divided into several orthogonal sub-bands dynamically allocated to individual users, which effectively eliminates intra-satellite interference~\cite{li2023online}.
		However, due to the antenna array and onboard power budget, the number of simultaneous narrow bandwidth spot beams that each satellite can form is limited. We denote this upper bound as $A_{max}$, representing the maximum number of users that can be concurrently served without mutual interference \cite{liu2024multi}. 
		To mitigate inter-satellite interference, we assume that all satellites within the same constellation are centrally coordinated and operate on disjoint frequency bands when covering overlapping regions~\cite{yang2024multi}, which aligns with practical frequency-reuse planning in large-scale satellite networks and ensures that co-channel interference between neighboring satellites can be safely ignored.
		The Signal-to-Noise Ratio (SNR) between user $u$ and satellite $s$ at time slot $n$ is given by:
		\begin{equation}
			\gamma_{u,s}(n)=\frac{P_{TX}G_{TX}L_{TX}G_{RX}L_{u,s}(n)}{\Theta F_{RX}W_{RF}}.
		\end{equation}
		
		The overall propagation loss between satellite $s$ and user $u$ at time slot $n$ is modeled as:
		\begin{equation}
			L_{u,s}(n)=L^f_{u,s}(n) L^r_{u,s}(n),
		\end{equation}  
		where $L^f_{u,s}(n)$ is the free‑space path loss at carrier frequency, and $L^r_{u,s}(n)$ is the rain attenuation. With the given carrier frequency $f$, the distance $d_{u,s}$ between satellite $s$ and user $u$, and the speed of light $c$, the free‑space loss is:
		\begin{equation}
			L^f_{u,s}(n)=\left(\frac c{4\pi d_{u,s}f}\right)^2.
			\label{free-space}
		\end{equation}
		Following ITU‑R P.838~\cite{recommendation2005838} and ITU-R P.618-14 ~\cite{recommendation2023618}, the rain attenuation is expressed as:
		\begin{equation}
			L^r_{u,s}(n)=10^{-\frac {\alpha R^\beta d_r} {10}},
				\label{rain}
		\end{equation}
		where $\alpha R^\beta$ is the attenuation per kilometer under a given rainfall rate, and the slant‑path length $d_r$ is calculated by:
		\begin{equation}
			\begin{aligned}
				&d_r=\begin{cases}\frac{(h_R-h_S)}{\sin\theta},&\theta\geq5^{\circ},\\\frac{2\left(h_R-h_S\right)}{\left(\sin^2\theta+\frac{2\left(h_R-h_S\right)}{R_e}\right)^{1/2}+\sin\theta},&\theta<5^{\circ},\end{cases}
			\end{aligned}
		\end{equation}
		where $h_S$ is the user’s antenna height, $h_R$is the rain‑height level, $\theta$ is the elevation angle, and $R_e$ is the effective Earth radius. And according to \cite{zhou2018channel}, the average rain height $h_R$ depends on latitude $\phi$:
		\begin{equation}
			\begin{aligned}
				&h_R=\begin{cases}5.0,&\phi\leq23^\mathrm{o},\\5.0-0.075\left(\phi-23\right),&\phi>23^\mathrm{o}.\end{cases}
			\end{aligned}
		\end{equation}  
		\subsubsection{Connection Model}
		The outage probability $P_{out}$ in communication is a key performance metric used to measure the reliability of a communication link, which is defined as the probability that an interruption will occur in a communication link during a given period. An outage occurs when the instantaneous SNR falls below a threshold $\gamma_{th}$: 
		\begin{equation}
			P_{out}=Pr\left(\gamma_{u,s}(n) \leq  \gamma_{th} \right).
		\end{equation}
		
		We assume that the parameters of transmitter and receiver remain constant during each time slot, hence, the propagation loss $L_{u,s}(n)$ is the only cause of outage:
		\begin{equation}
			P_{out}=Pr\left(L_{u,s}(n) \geq  L_{th} \right).
		\end{equation}
		
		According to \eqref{free-space} and \eqref{rain}, the propagation loss mainly depends on the distance $d_{u,s}$ and slant‑path length $d_r$. As the elevation angles $\theta_{u,s}(n)$ decrease, the distance $d_{u,s}$ and slant‑path length $d_r$ increase, which results in higher total attenuation and elevated outage probability:
		\begin{equation}
			P_{out}=Pr\left(\theta_{u,s}(n) \geq  \theta_{min} \right).
		\end{equation}
		Therefore, satellite communication is typically restricted to elevation angles above a minimum threshold to enhance link reliability. Based on the geometric relationship, $\theta_{u,s}(n)$ can be calculated as follows~\cite{gongora2022link}:
		\begin{equation}
			{\theta}_{s,u}\!(n)\!=\!\arctan\!\left(\!\frac{\cos\psi_{s,u,n}\cos\phi_{u,n}\!-\!(r_{G,O}/r_{S,O})}{\sqrt{1-\cos^2\psi\cos^2\phi_{u,n}}}\right),
		\end{equation}
		where $\psi_{s,u,n}$ denotes the longitudinal separation between user $u$ and satellite $s$ at time slot $n$, while $\phi_{u,n}$ is the latitude of user $u$ at time slot $n$. The parameters $r_{G,O}$ and $r_{S,O}$ represent the radial distances from the Earth’s center to the user and to the satellite, respectively.
		
		We introduce the binary scheduling variable $a_{u,s}(n)$, where if user $u$ lies within the coverage of satellite $s$ and is served by it at time $n$, $a_{u,s}(n) = 1$; otherwise, $a_{u,s}(n) = 0$. Although a user may fall within the footprints of multiple satellites, it may be assigned to at most one satellite in any time slot:
		\begin{equation}
			\sum\nolimits_{s=1}^{S}a_{u,s}(n)=1 , \forall u \in \mathcal{U}, \forall n \in \mathcal{N}.
		\end{equation}
		Meanwhile, each satellite forms a finite number of narrowband steerable beams that dynamically track their target users \cite{liu2024multi}. Hence, the number of simultaneously served users cannot exceed the number of available beams:
		\begin{equation}
			\sum\nolimits_{u=1}^{U}a_{u,s}(n)\leq A_{max}, \forall s \in \mathcal{S}, \forall n \in \mathcal{N}.
		\end{equation}
		Finally, to guarantee reliable link quality, any scheduled user–satellite pair must satisfy a minimum elevation‐angle requirement. This is enforced via:
		\begin{equation}
			(1-a_{u,s}(n)) \Gamma+{\theta}_{s,u}(n) \geq \theta_{min}, 
		\end{equation} 
		$\theta_{min}$ is the threshold elevation angle, $\Gamma$ is a large constant that renders the constraint inactive whenever $a_{u,s}(n) = 0$.

		\subsubsection{Communication Rate}
		To accommodate the heterogeneous service requirements of static GUs and mobile FUs, while mitigating the inefficiencies associated with fixed and discrete resource block allocation, we propose a dynamic resource allocation framework that flexibly and jointly allocates orthogonal frequency channels to both types of users.
		 Specifically, a key component of this approach is the bandwidth allocation coefficient $b_{u,s}(n) \in [0,1]$, representing the fraction of satellite $s$’s total available bandwidth allocated to user $u$~\cite{vu2021dynamic},~\cite{wang2022bandwidth}. Under this scheme, the set of orthogonal subcarrier channels is subject to:
		\begin{equation}
			\sum\nolimits_{u=1}^{U}b_{u,s}(n)\leq 1, \forall s \in \mathcal{S}, \forall n \in \mathcal{N},
		\end{equation}
		thereby ensuring that each user receives a continuous, proportionate share of bandwidth tailored to its instantaneous demand. By adjusting $b_{u,s}(n)$ in real time, the framework maximizes spectral efficiency and alleviates resource underutilization, while still guaranteeing that all scheduled users maintain the minimum QoS. Hence, the communication rate can be expressed as:
		\begin{equation}
			R_{u,s}(n)=b_{u,s}(n) W\mathrm{log}_2(1+\gamma_{u,s}(n)),
		\end{equation}
		 where, $W$ is the total bandwidth of satellite $s$.
		
		\vspace{-1mm}
		\subsection{Service Migration Model}
		To realistically model service deployment and migration in satellite networks, we consider a three-layer framework~\cite{chen2024spaceedge} for service migration, which can be divided into: base layer, application layer, and instance layer. 
		\subsubsection{Base Layer}The base layer includes core system components such as the operating system and kernel, which are independent of specific services and can be shared across all applications. This layer is typically pre-installed on satellite servers and thus excluded from the migration payload. 
		\subsubsection{Application Layer}The application layer contains service-specific but user-agnostic data, such as libraries, configuration files, or shared content, which is deployed on demand and reused by multiple user instances. This reduces transmission overhead, as the application layer does not need to be repeatedly transferred for each session. 
		\subsubsection{Instance Layer}The instance layer stores user-specific runtime states, including volatile memory, CPU context, and execution history. As it reflects the personalized, real-time execution of a service, this layer must be migrated to ensure seamless service continuity. 
		
		Service migration is modeled as a stateful process that involves transferring the user-specific execution state of an instance to a new host~\cite{wang2018survey}. 
		According to the adopted three-layer framework, only the instance layer, which captures volatile runtime information such as CPU state, memory content, and intermediate results, needs to be migrated, while the base and application layers are either pre-installed or deployed on demand. 
		Specifically, the base layer is pre-installed on all satellite servers~\cite{chen2024spaceedge,li2023online}, incurring a one-time storage cost at the commissioning stage that can be regarded as a long-term overhead beyond online scheduling.  
		For the application layer, we assume that it is fully deployed on all satellite servers at the initial stage, and subsequently updated in a periodic manner. 
		Recent studies \cite{machen2017live} have shown that the application layer can be distributed independently of a running service and even pre-distributed through demand-aware caching, allowing updates to be performed in the background without incurring additional real-time migration burden. 
		Accordingly, we define a fixed initialization cost $\tau_{\text{app}}$ to represent the periodic expense of application updates. 
		This abstraction allows us to capture the long-term effect of base layer storage and the periodic nature of application deployment, while keeping the focus on the dynamic migration and resource allocation of the instance layer, which is the dominant source of overhead during frequent handovers.

\begin{figure}[t]
			\centering
			\includegraphics[width=0.45\textwidth]{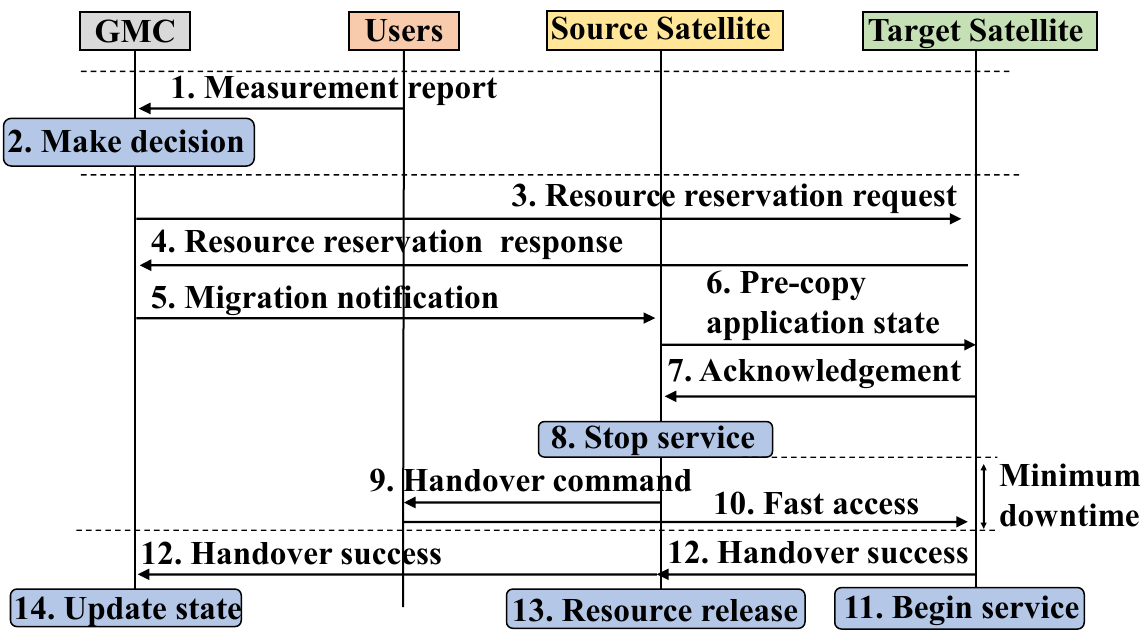}\vspace{-2mm}
			\caption{Service migration procedures.}\vspace{-6mm}
			\label{fig:2}
		\end{figure}

		We employ a live migration strategy with pre-transmission to minimize service interruptions during the migration process. 
		Nevertheless, due to inevitable synchronization delays and final state handoff, a non-zero service outage period may still occur~\cite{li2023online}. 
		Based on the illustrated procedures in Fig.~\ref{fig:2}, when a user $u$ moves from source satellite $s_1$ to target satellite $s_2$, computing resources $f_{u,s_2}(n)$ must be pre-allocated on $s_2$. 
		Once confirmed, the runtime state is pre-copied to $s_2$, and the service is briefly paused for final synchronization. 
		The target satellite then quickly resumes the service, minimizing interruption. 
		Resources on the source satellite are subsequently released. 
		To meet service quality, the allocated resources must satisfy $f_{u,s}(n)\geq \zeta_u$, where $\zeta_u$ is the user’s minimum computing requirement.

		Moreover, service migration introduces additional operational overhead, such as bandwidth consumption and satellite processing cost. To jointly account for both factors, we define a unified migration cost term $\beta_u^1$, and introduce a binary indicator variable $C_{u,s}^1 (n)$ to represent whether user $u$' service is migrated at time slot $n$:	
		\begin{equation}
			C_{u,s}^1 (n)=a_{u,s}(n)\oplus a_{u,s}(n-1),
		\end{equation}
		where $\oplus$ denotes the XOR operation. A value of 1 indicates that the user was reassigned to a different satellite between time slots, triggering a migration event.

		\vspace{-1mm}
		\subsection{Traffic Model}
		\vspace{-1mm}
		To effectively capture the heterogeneous service demands in satellite networks, we consider two distinct user types: GUs and FUs. GUs are stationary and maintain persistent service demands throughout the entire period, typically generating high-volume, delay-sensitive traffic. In contrast, FUs exhibit mobility along predefined flight trajectories and initiate service requests only during the stable cruising phase, typically with lighter workloads. Despite this binary classification, users within each group may still exhibit significant variations in their service characteristics. 
		To model such diversity, each user $u \in \mathcal{U}=\mathcal{U}_1 \cup \mathcal{U}_2$ is represented by a six-tuple $	(d_u,t_u,\zeta_u, \lambda_u, \beta_u^1,\beta_u^2),$
		where $d_u$ is the average packet size, $t_u$ is the maximum tolerable delay, $\zeta_u$ denotes the minimum required computing resource, and $\lambda_u$ is the packet arrival rate, assumed to follow a Poisson distribution. The parameters $\beta_u^1$ and $\beta_u^2$ quantify the cost of service migration and the corresponding service utility, respectively. All parameters are user-specific and capture intra-group heterogeneity.
		
		Given the high traffic demand and limited service capacity of satellites, especially when concurrently supporting a large number of users, it becomes essential to model the contention for communication and computational resources. To this end, we adopt a queuing-based approach to manage packet transmissions from each user's application. 
		Specifically, we define $\mu_u$ as the number of packets that can be transmitted and processed per unit of time, expressed as: $\mu_u=R_{u,s}\delta/d_u.$
		Each satellite may host multiple user applications, and each user maintains a separate uplink queue that operates under a First-Come-First-Served (FCFS) policy. Considering the long and variable propagation delays typical in satellite links, we approximate the service time as exponentially distributed, thereby modeling the system as an M/M/1 queue \cite{huang2024aoi}. This abstraction captures the random arrival and service processes while offering tractable analysis.
		
		The sojourn time $W^{total}_{u,s,i}$ of the $i$-th packet for user $u$ consists of the queuing delay $T^{que}_{u,s,i}$ and transmission time $T^{tran}_{u,s,i}$, satisfying: $	W^{total}_{u,s,i}=T^{que}_{u,s,i}+T^{tran}_{u,s,i}.$  
		To guarantee QoS, this total delay must not exceed the user-specific delay threshold $t_u$, i.e., $W^{total}_{u,s,i} \leq t_u.$
		In the proposed FCFS queuing system, when packet $i$ arrives and finds $k>1$ packets already in the system, its waiting time consists of two components: the residual service time of the packet currently in service and the full service times of the $k-1$ packets ahead in the queue. As a result, the waiting time follows a $k$th-order Erlang distribution. Hence, the total sojourn time follows a $(k+1)$th-order Erlang distribution. Let $W(x)$ denote the cumulative distribution function (CDF) of the sojourn time. The unconditional probability that the sojourn time is no greater than $x$ is given by marginalizing over all possible queue lengths $k$ at the time of arrival:
		\begin{equation}
			\small
			\begin{aligned}
				W(x)& =P\left\{0<W\leqslant x\right\} \\
				&=\sum\nolimits_{k=0}^{\infty}\lim_{n\to\infty}P\left\{N_{\tau_{n}^{-}}=k\right\}P\left\{0<W\leqslant x|N_{\tau_{n}}=k\right\}. \\				
			\end{aligned}
		\end{equation}
		Assuming M/M/1 queue with arrival rate $\lambda$ , service rate $\mu$, and system utilization $\rho=\lambda/\mu$, the CDF simplifies to:
		\begin{equation}
			\begin{aligned}
				W(x)&=(1-\rho)\int_{0^{+}}^{x}\mu\sum\nolimits_{k=0}^{\infty}\frac{(\mu y\rho)^{k}}{k!}\mathrm{e}^{-\mu y}\mathrm{d}y \\
				&=1-e^{-\mu(1-\rho)x}.				
			\end{aligned}
		\end{equation}
		This expression indicates that the sojourn time in the M/M/1 queue under FCFS discipline follows an exponential distribution with parameter $\mu(1-\rho)$, reflecting the effect of queue congestion and processing capacity on the latency performance. Accordingly, we define $C_{u,s}^2(n)$ as the number of bits successfully transmitted for user $u$, accounting for the delay constraint: $	C_{u,s}^2(t)=\lambda_u \delta d_u Pr(W \leq t_u),$
		which reflects the delay-aware service capability in the satellite network.

		\vspace{-1mm}
		\section{Problem Formulation}
		\label{pro_for}
		\vspace{-1mm}
		In this section, we formulate the joint resource allocation and service migration problem in the satellite network. 
		Due to the stochastic and heterogeneous nature of service requests, we aim to maximize the long-term system utility, defined as the cumulative utility accrued over the entire period. At each time slot $n$, the instantaneous utility derived from serving user $u$ by satellite $s$ is expressed as:
		\begin{equation}
			Utility_{u,s}(n)=a_{u,s}(n)[\beta_u^1C_{u,s}^1 (n)+\beta_u^2C_{u,s}^2 (n)].
		\end{equation}
		
		The optimization objective is to maximize the total system utility over the time period $N$ by jointly optimizing the user association $A=\{a_{u,s}(n)\}$, bandwidth allocation $B=\{b_{u,s}(n)\}$, and computing resource allocation $F=\{f_{u,s}(n)\}$:
		\begin{subequations}
			\begin{align}
				&\max_{A,B,F} \sum_{n=1}^{N}\sum_{u=1}^{U}\sum_{s=1}^{S}a_{u,s}(n)[\beta_u^1C_{u,s}^1 (n)+\beta_u^2C_{u,s}^2 (n)]  \label{proa}\\
				&\quad\text{s.t.}\quad\sum\nolimits_{s=1}^{S}a_{u,s}(n)=1 , \forall u \in \mathcal{U},  \label{prob}\\
				&\quad\qquad\sum\nolimits_{u=1}^{U}a_{u,s}(n)\leq A_{max}, \forall s \in \mathcal{S},  \label{proc}\\
				&\quad\qquad\sum\nolimits_{u=1}^{U}a_{u,s}(n)b_{u,s}(n)\leq B_{max}, \forall s \in \mathcal{S},   \label{prod}\\
				&\quad\qquad \sum\nolimits_{u=1}^{U}a_{u,s}(n)f_{u,s}(n)\leq F_{max}, \forall s \in \mathcal{S},   \label{proe}\\	
				&\quad\qquad f_{u,s}(n)\geq \zeta_u, \forall u \in \mathcal{U}, \forall s \in \mathcal{S},  \label{prof} \\
				&\quad\qquad 	(1-a_{u,s}(n)) \Gamma+{\theta}_{s,u}(n) \geq \theta_{min},  \label{prog} \\
				&\quad\qquad a_{u,s}(n)\in\{0,1\},b_{u,s}(n)\geq0,f_{u,s}(n)\geq0.  \label{proh}
			\end{align}
		\end{subequations}
		
		The constraints in this problem are detailed as follows: Constraint \eqref{prob} ensures that each user is associated with at most one satellite per time slot, Constraint \eqref{proc} restricts the maximum number of users a satellite can serve simultaneously, Constraint \eqref{prod} ensures the total allocated bandwidth does not exceed satellite capacity, Constraint \eqref{proe} limits the computing workload a satellite can handle, Constraint \eqref{prof} ensures each active user receives at least its baseline computational requirement, and Constraint \eqref{prog} enforces a minimum elevation angle $\theta_{min}$ between the satellite and the user to maintain reliable communication. All constraints hold per time slot $n \in \mathcal{N}$. This problem captures the key resource constraints in satellite networks and the dynamic service requirements of heterogeneous users. Due to its combinatorial and nonlinear nature, it is computationally challenging and motivates the development of efficient learning-based optimization methods. 
		
		\begin{figure*}[!t]
			\centering
			\includegraphics[width=1\textwidth]{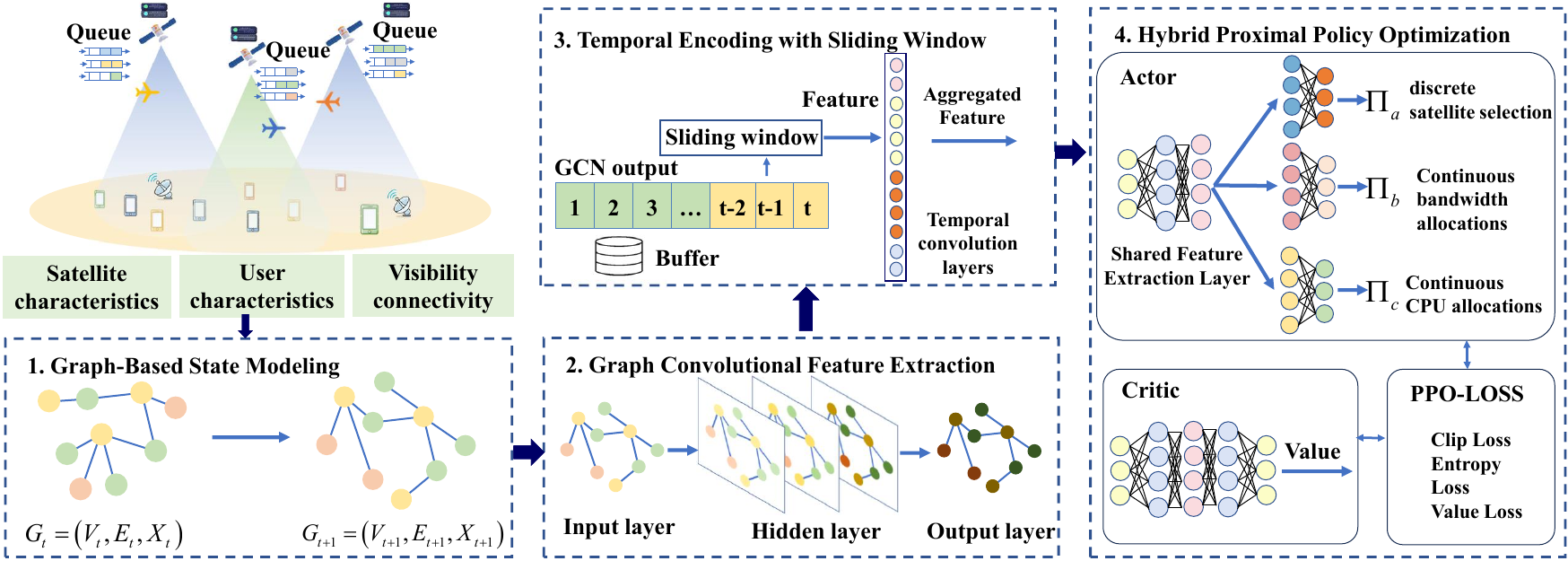}\vspace{-2mm}
			\caption{Graph-Aware Temporal Encoder for Reinforcement Learning (GATE-RL)}\vspace{-6mm}
			\label{fig:3}
		\end{figure*}
		\vspace{-1mm}
		\section{Graph-Aware Temporal Encoder for Reinforcement Learning}
		\label{scheme}

		The optimization problem formulated by (20) exhibits a non-convex structure and is proven to be NP-hard, rendering traditional optimization techniques inadequate~\cite{du2024fmWCM}. 
		To address this, we reformulate the problem as an MDP and adopt RL as a scalable and effective approach for discovering high-quality policies in complex environments.
		The application of RL in satellite-user systems entails significant spatial and temporal complexities. Temporally, abrupt fluctuations in satellite resources and the sporadic nature of service requests, particularly from FUs, render single-time step observations insufficient. Spatially, dynamic connectivity patterns emerge due to the mobility of satellites and FUs, as well as the heterogeneous characteristics of different users. These factors necessitate learning frameworks capable of capturing both temporal dependencies and graph-structured spatial relationships.
		
		To address the aforementioned spatiotemporal challenges, we design a RL framework, GATE-RL, which integrates Graph Neural Networks (GNNs) and temporal convolution within the HPPO backbone. 
		As illustrated in Fig.~\ref{fig:3}, the proposed architecture first constructs a dynamic graph representation of the satellite-user system at each time step, where nodes correspond to satellites and users, and edges capture time-varying connectivity. A GNN is employed to extract spatially-aware representations of the system state. To capture short-term temporal dependencies and reduce scheduling instability, we introduce a temporal encoding module based on 1D convolution over a fixed-length historical window of extracted graph features. Both the actor and critic networks then utilize the resulting temporal-spatial representation to inform decision-making. Furthermore, we adopt a hybrid action space formulation, where discrete actions are used for service deployment decisions, and continuous actions are used for fine-grained resource allocation.
		\vspace{-3mm}
		\subsection{Graph-Based State Modeling}
		To systematically capture the spatial structure and resource distribution in the satellite-user system, we model the environment as a dynamic graph. At each time slot $n$, the system state is represented as a graph $G_{n}=\left ( V_n, E_n,X_n \right )$, where:
		\begin{itemize}
			\item $V_n$ denotes the set of nodes comprising both satellite, GUs, and FUs nodes.
			\item $E_n$ represents the set of edges defined by satellite-user visibility relationships. An edge $(u,s) \in E_n$ exists if and only if user $u$ is within the communication coverage area of satellite $s$ at time slot $n$. 
			\item $X_n \in \mathbb{R}^{\lvert V_{n} \rvert \times d }$ is the node feature matrix, where each row corresponds to the feature vector of a specific node, and $d$ is the feature dimension.
		\end{itemize} 
		
		The feature vectors for different types of nodes are designed to encapsulate task-relevant system information. For satellite nodes, the features include coordinates, remaining communication bandwidth, available computational resources, and the number of remaining access slots. These attributes jointly characterize the satellite’s service capability and spatial location at time $n$. For user nodes, the features consist of coordinates, task priority, task arrival rate, and the identifier of the satellite that hosted the user’s task during the previous scheduling period. This design captures both the current service demand and the temporal continuity of service provisioning.
		To effectively utilize these features, the system state at each time step is modeled as a graph structure $G_n$, where each node corresponds to a satellite or user, and edges represent visibility-based connections. This graph formulation preserves both local state information and the relational dependencies among nodes. It serves as the input to GNN layers, which are employed to extract spatially correlated features that are essential for informed and coordinated decision-making.
		\vspace{-3mm}
		\subsection{Graph Convolutional Feature Extraction}
		To effectively capture the spatial dependencies in the satellite-user system and adapt to the periodic orbital dynamics, we model the environment as a dynamic graph $G_n$ at each time slot $n$, where both node features and adjacency are time-varying. A GCN is employed as the core feature extractor in both the actor and critic networks. 
		We begin with the spectral definition of graph convolution. The normalized graph Laplacian is:
		\begin{equation}
			L_n = I - D_n^{-\frac12}\,A_n\,D_n^{-\frac12},
		\end{equation}
		 where, $A_n\in\mathbb{R}^{|V_n|\times|V_n|}$ and $D_n$ are the adjacency matrix and degree matrix of the graph $G_{n}$, respectively.
		 Its eigendecomposition is $L_n = U_n \Lambda_n U_n^\top$ yields the orthonormal eigenvectors $U_n$ and diagonal eigenvalues $\Lambda_n=\mathrm{diag}(\lambda_1,\dots,\lambda_{|V_n|})$.  A spectral filter $g_\theta$ acting on node features $X_n\in\mathbb{R}^{|V_n|\times d}$ is then given by:
		\begin{equation}
			g_\theta \star X_n = U_n\,g_\theta(\Lambda_n)\,U_n^\top\,X_n.
			\label{filter}
		\end{equation}
		
		In practice, computing the full eigendecomposition is computationally prohibitive for large-scale and time-varying graphs. Therefore, we adopt a first-order approximation of \eqref{filter}, which replaces the spectral operation with a localized aggregation scheme based on the symmetrically normalized adjacency matrix \cite{cai2024graphic}:
		\begin{equation}
			\hat A_n = D_n^{-\frac12}\,(A_n + I)\,D_n^{-\frac12},
		\end{equation}
		where $A_n$ is the raw adjacency matrix defined by satellite-user visibility at time $n$, $I$ is the identity matrix adding self-loops to all nodes, $D_n$ is the degree matrix corresponding to $A_n+I$.

		While the GCN architecture follows standard formulations, its application to the satellite-user system introduces several critical adaptations. First, the adjacency matrix $\hat{A}_n$ is recomputed at each time step $n$ based on real-time satellite-user visibility relationships, enabling the model to adapt to the rapidly changing satellite networks topology. Second, the feature matrix $X_n$ contains distinct attribute sets for satellites, GUs, and FUs, enabling the model to capture the heterogeneous nature of different node types and their specific roles in the system. And the node features explicitly encode orbital dynamics through 3D coordinates and remaining access slots, allowing the GCN to learn spatial relationships that respect orbital mechanical constraints. This dynamic and heterogeneous graph formulation enables the GCN to generalize across rapidly evolving topologies, providing a principled approach to capturing orbital periodicity and spatial correlations in non-stationary satellite networks.
		Hence, at each time slot $n$, given the graph structure defined by the normalized adjacency matrix with self-loops $\hat{A_n} \in \mathbb{R}^{\lvert V_{n} \rvert \times \lvert V_{n} \rvert }$, and the node feature matrix $X_n \in \mathbb{R}^{\lvert V_{n} \rvert \times d }$, the GCN operates on these dynamically-updated graphs, performing localized aggregation. Specifically, we adopt a two-layer GCN, formulated as:
		\begin{equation}
			H_n=\mathrm{ReLU}\left(\hat{A}_n\cdot\mathrm{ReLU}\left(\hat{A}_n\cdot X_nW_1\right)W_2\right),
		\end{equation}
		where $H_n$ captures spatially-aware node embeddings at time $n$. By updating $X_n$ and $A_n$ at each time step according to satellite positions and visibility, the GCN effectively adapts to the periodic and dynamic nature of satellite-user connectivity.
		$W_1 \in \mathbb{R}^{d \times h_1 }$ and $W_2 \in \mathbb{R}^{h_1 \times h }$ are trainable weight matrices associated with the first and second GCN layers, respectively. $\mathrm{ReLU}(\cdot)$ is the rectified linear unit activation function applied element-wise.
		
		The first layer of the GCN propagates information from each node’s immediate neighbors, including itself, by transforming raw features $X_n$ into an intermediate representation $H_n^{(1)}$. The second layer further refines these representations $H_n^{(1)}$ by repeating the neighborhood aggregation, effectively enabling two-hop message passing. This two-layer structure strikes a balance between expressive capacity and computational efficiency, enabling the model to learn spatial feature embeddings that capture both local node states and neighboring interactions, which are essential for informed decision-making in the satellite-user system.
		
		By integrating these graph-aware representations $H_n$, the actor and critic networks gain access to spatially correlated insights essential for downstream decision-making, enhancing both policy learning and value estimation.
		\vspace{-2mm}
		\subsection{Temporal Encoding via Sliding Window}
		While GCN effectively captures spatial correlations at each individual time step, the highly dynamic nature of satellite–user systems necessitates the modeling of short-term temporal dependencies. To this end, we introduce a fixed‐length sliding window of size $W$, which retains the most recent $W$ GCN‐encoded graph representations:
		\begin{equation}
			\{H_{n-W+1},H_{n-W+2},\,\dots,\,H_n\},\quad H_\tau\in\mathbb{R}^{|V_n|\times h}.
		\end{equation}
		We flatten each $H_\tau$ to $\mathrm{vec}(H_\tau)\in\mathbb{R}^{|V_n|\,h}$ and stack them into
		\begin{equation}
			\mathcal{S}_n = \bigl[\mathrm{vec}(H_{n-W+1}),\dots,\mathrm{vec}(H_n)\bigr]\in\mathbb{R}^{(|V_n|h) \times W}.
		\end{equation}
		To extract temporal features from $\mathcal{S}_n$, we apply a one-dimensional convolutional layers denote by $\mathrm{Conv1}$ with kernel size $k \geq W$ and output channels $C$. This is followed by Layer Normalization $LN(\cdot)$ and ReLU activations, yielding:
		\begin{equation}	
		Y= \mathrm{ReLU}\!\bigl(\mathrm{LN}(\mathrm{Conv1}(\mathcal{S}_n))\bigr).
		\end{equation}
		The output is then compressed by removing the temporal dimension, resulting in a fixed-size temporal feature code $Z_n \in\mathbb{R}^{C} $.
		This architecture enables the model to learn rich and hierarchical temporal abstractions over recent graph embeddings. By doing so, it mitigates the instability of myopic, single-step decisions and captures short-term trends in user demand patterns and satellite resource availability. The extracted temporal code $Z_n$ is subsequently fed into both the actor and critic networks to support robust policy optimization and value estimation.
		\vspace{-1mm}
		\subsection{Markov Decision Process Formulation}

		To solve the joint service placement and resource allocation problem in satellite networks, we formulate it as an MDP $\mathcal{M}=(\mathcal{S},\mathcal{A},P,r,\gamma)$.  At each discrete time slot \(n\), a centralized intelligent agent (the Ground Management Controller, GMC) observes the current network state \(s_n\), selects a hybrid action \(a_n\), and receives an immediate reward \(r_n\) before the system transitions to \(s_{n+1}\).  The objective of the agent is to learn a stochastic policy \(\pi_\theta(a_n\!\mid s_n)\) that maximizes the expected cumulative discounted reward \(\mathbb{E}_{\pi}[\sum_{n=0}^\infty \gamma^n\,r_n]\).
		
		\subsubsection{State Space \(\mathcal{S}\)}
		
		At each time \(n\), the state \(s_n \in \mathcal{S}\) is defined as a temporally-augmented graph representation $Z_n$, which integrates both current and recent graph-structured system information.
		This includes satellite–user connectivity, ongoing service assignments, and resource availability dynamics. As described in earlier sections, this state is computed using GCNs followed by a Temporal Convolutional Network, enabling the agent to reason over both spatial and short-term temporal dependencies.
		
		\subsubsection{Action Space \(\mathcal{A}\)}
		
		The agent determines both service placement and resource allocation for each user at each time slot. The hybrid action space is defined as:
		\begin{equation}
			\mathcal{A}_n = \left\{ \left( a_{u,n}, b_{u,n}, f_{u,n} \right) \,\middle|\, u \in \mathcal{U} \right\},
		\end{equation}
		where for each user \(u\), \(a_{u,n} \in \{1, \dots, N_{\text{sat}}\}\) is a discrete action denoting the selected satellite for service deployment; \(b_{u,n} \in (0, 1]\) is a continuous variable representing the communication bandwidth allocation ratio; \(f_{u,n} \in (0, 1]\) is a continuous variable representing the computational resource allocation ratio.
		To ensure bounded and stable continuous outputs, the variables \(b_{u,n}\) and \(f_{u,n}\) are sampled from parameterized Beta distributions, which inherently constrain the values to the interval \((0, 1]\), thereby offering improved stability during training compared to truncated Gaussian distributions.

		\subsubsection{Transition Probability \(\mathcal{P}\)}
		
		The state transition probability \(\mathcal{P}(s_{n+1} \,|\, s_n, a_n)\) captures the environment's stochastic evolution. It depends on: random task arrivals and heterogeneous service demands from users; satellite mobility and changing visibility conditions; resource depletion or release resulting from allocation and migration decisions..
		Since the exact transition dynamics are unknown and potentially non-stationary, they are implicitly learned by the agent through trial-and-error interaction with the environment during training.
		
		\subsubsection{Reward Function \(r\)}
		The immediate reward at time slot \(n\) is defined as:
		\begin{equation}
			\small
			r_n = \sum_{u \in \mathcal{U}}  \beta_u^2 C_{u,n}^2   - \lambda_{\text{pen}}  \left(\sum_{u \in \mathcal{U}}\beta_u^1 C_{u,n}^1  + \text{Penalty}(a_n) \right),
		\end{equation}
		where \(C_{u,n}^1\) indicates whether user $u$' service is migrated at time slot $n$, \(C_{u,n}^2\) denote task utility terms for user\(u\). 
		$\text{Penalty}(a_n)$ is a constraint function that imposes penalties when system limits are breached, including bandwidth capacity, computational resource capacity, and connection capacity constraints.
		$\lambda_{\text{pen}}$ is a regularization coefficient that controls the impact of constraint violations on the reward.
		Thus, the reward not only balances migration and resource utilization but also implicitly reflects delay and energy consumption.

		\vspace{-1mm}
		\subsection{Hybrid Proximal Policy Optimization}
		To effectively optimize the hybrid discrete-continuous action space arising in satellite-user system scheduling, we propose an HPPO framework. This framework extends the standard PPO algorithm by integrating graph convolutional feature extraction, temporal convolution encoding, and multi-head policy output, thereby supporting mixed action modalities while leveraging spatial-temporal graph structures.
		
		In this formulation, the agent first extracts a spatial-temporal representation for each state \( s_n \) using the GATE. This encoder combines GCNs and Temporal Convolutional Networks to produce a latent embedding \( z_n \). This shared embedding is subsequently fed into both the actor and critic networks.

		The actor network employs a multi-head structure to generate hybrid actions from the shared temporal-spatial embedding \( z_n \) of the GATE.  Specifically, for each user \( u \), the shared embedding is first processed by a common base network to extract a high-level latent representation. This latent representation is then fed in parallel into three separate output heads.
		The discrete head for satellite selection outputs a categorical distribution \(\pi_\theta^{A}\) over available satellites.
		For the continuous actions of CPU and bandwidth allocation, which represent bounded resource ratios, the model parameterizes Beta distributions \(\pi_\theta^{B}\) and \(\pi_\theta^{F}\). This choice is principled: the Beta distribution is natively defined on the \((0,1]\) interval, providing an inherent match to the physical constraints of resource allocation problems. This inherent boundedness promotes training stability by eliminating the pathological behavior and bias associated with clipping the outputs of unbounded distributions like the Gaussian. Each continuous head produces the concentration parameters for its Beta distribution via dedicated sub-networks, enabling the policy to learn diverse, non-uniform allocation strategies within the valid action space. This multi-head, per-user design facilitates specialized decision-making while leveraging a common feature foundation.

		To train the actor, we adopt a hybrid form of the clipped surrogate PPO objective, which applies separately to each action type while sharing advantage estimates:
\begin{multline}
	\vspace{-2mm}
	L^{\text{Hybrid-PPO}}(\theta) = 
	\mathbb{E}_{n} \left[ \sum_{k \in \{A, B, F\}} 
	\min \left( r_{n}^{(k)}(\theta) \hat{A}_n, \right. \right. \\
	\left. \left. \mathrm{clip}\left(r_{n}^{(k)}(\theta), 1-\epsilon, 1+\epsilon\right) \hat{A}_n \right) \right],
\end{multline}
		where \( r_{n}^{(k)}(\theta) = \frac{\pi_\theta^{(k)}(a_{n}^{(k)} \,|\, s_n)}{\pi_{\theta_{\text{old}}}^{(k)}(a_{n}^{(k)} \,|\, s_n)} \) denotes the probability ratio under the new and old policies for action type \( k \), and \(\hat{A}_n\) is the advantage estimate computed using Generalized Advantage Estimation (GAE). The GAE is calculated as follows:
		\begin{equation}
			\vspace{-1mm}
			\hat{A}_n = \sum\nolimits_{l=0}^{\infty} (\gamma \lambda)^l \delta_{n+l},
		\end{equation}
		where the temporal difference error is:
		\begin{equation}
			\vspace{-1mm}
			\delta_n = r_n + \gamma V_\phi(s_{n+1}) - V_\phi(s_n),
		\end{equation}
		with \( V_\phi(s_n) \) being the critic’s estimation of the value function.
		
		The critic network is optimized by minimizing the standard value function regression loss:
		\begin{equation}
			L^{\text{Critic}}(\phi) = \frac{1}{2} \mathbb{E}_{n} \left[ \left( V_\phi(s_n) - \hat{V}_n \right)^2 \right],
		\end{equation}
		where the target value estimate is given by \(\hat{V}_n = \hat{A}_n + V_\phi(s_n)\).
		
		By employing a shared encoder between the actor and critic, the proposed HPPO enhances both learning stability and sample efficiency under structured graph-based states. Moreover, the multi-head actor design enables efficient optimization over heterogeneous actions, naturally accommodating the mixed discrete–continuous decision structure of the scheduling problem. This unified framework preserves the benefits of proximal policy optimization, such as monotonic policy updates and stable training, while explicitly leveraging spatial–temporal correlations critical to satellite–user coordination. The algorithm is detailed in Algorithm~\ref{alg:HPPO}.
		\vspace{-1mm}
		\subsection{Computational Complexity Analysis}
		 We focus on the forward propagation complexity, as the online runtime depends mainly on the actor network; the critic is used only during training and does not affect real-time inference. The inference phase of the proposed GATE-HPPO framework consists of three components:
		 Given a graph with $|E|$ edges, feature dimension $d$, and $C$ input channels, an $n_g$-layer GCN requires $\mathcal{O}(n_g |E| d C)$ operations for forward propagation \cite{dong2021intelligent}. 
		 Here, $|E|$ corresponds to the number of visible satellite-user links at each time slot, reflecting the connectivity density of the dynamic constellation graph. The parameter $d$ denotes the dimension of the extracted feature vector for each node, which includes satellite–user connectivity, ongoing service assignments, and resource availability dynamics. The parameter $C$ represents the number of input channels, which is equivalent to the number of hidden features in a GCN layer. Finally, $n_g$ indicates the number of stacked GCN layers. Since $n_g$, $d$, and $C$ are hyperparameters independent of the graph size, the overall complexity remains linear in the number of edges $|E|$, which is scalable for large satellite networks.
		 The GCN outputs from the latest $T$ time steps are processed by a temporal convolutional layer with $c_t$ channels and kernel size $k_t$, incurring a complexity of $O(T \, d_g \, c_t \, k_t)$, where $d_g$ denotes the GCN output dimension. 
		 The actor network has $L_a$ fully-connected layers and $m$ output heads. Let $n_i^{(l)}$ and $n_o^{(l)}$ denote the input and output dimensions of layer $l$. 
		 Then the forward complexity is $O\Big(m \sum_{l=1}^{L_a} n_i^{(l)} n_o^{(l)} \Big)$, where $n_i^{(1)} = n_{\text{in}}$ is the input from the temporal convolution.
		 Combining the above, the total forward-propagation complexity is $O(n_g |E| F C) + O(T \, d_g \, c_t \, k_t) + O\Big(m \sum_{l=1}^{L_a} n_i^{(l)} n_o^{(l)} \Big)$, which scales linearly with $|E|$ while other terms are constant given fixed hyperparameters.

		\begin{algorithm}[htbp]
			\caption{ Graph-Aware Temporal Encoder for Hybrid Proximal Policy Optimization (GATE-HPPO)}
			\label{alg:HPPO}
			\begin{algorithmic}[1]
				\Statex \textbf{Input:} GATE-HPPO model, Simulation environment
				\Statex \textbf{Output:} GATE-HPPO model
				\State \textbf{Initialize:} GATE parameters $u$, Actor parameters $\theta$, Critic parameters $\phi$, Replay Buffer $\mathcal{D}$, Time window size $W$
				\For{each iteration}
				\For{each environment step $n$}
				\State \textbf{Collect Environment State:} Observe raw features 
				\Statex \quad \quad \quad \quad $\{\theta_n, \text{resource}_n, \text{users}_n, \text{schedule}_{n-1}\}$
				\State \textbf{Graph Construction:} Build $G_n = (V_n, E_n, X_n)$ 
				\State \textbf{Graph Convolutional Embedding:} Extract node 
				\Statex \quad \quad  \quad \quad embeddings $H_n = \mathrm{GCN}(G_n)$
				\State \textbf{Temporal Feature Aggregation:} Maintain sliding 
				\Statex \quad \quad \quad \quad window $\{H_{n-W+1}, \dots, H_n\}$. Encode spatial-
				\Statex \quad \quad \quad \quad temporal feature $z_n$
				\State \textbf{Action Sampling:}
				\State  \quad Sample discrete satellite $a_n^A \sim \pi_\theta^A(\cdot | z_n)$
				\State  \quad Sample continuous bandwidth $a_n^B \sim \pi_\theta^B(\cdot | z_n)$
				\State  \quad Sample continuous CPU $a_n^F \sim \pi_\theta^F(\cdot | z_n)$
				\State \textbf{Environment Step:} Execute $a_n = \{a_n^A, a_n^B, a_n^F\}$, 
				\Statex \quad \quad \quad \quad observe reward $r_n$ and next state $s_{n+1}$
				\State \textbf{Store:} Add $(s_n, a_n, r_n, s_{n+1})$ to $\mathcal{D}$
				\EndFor
				\State \textbf{Compute Advantage Estimates:} $\hat{A}_n$ using GAE
				\For{each PPO update epoch}
				\State \textbf{Sample Mini-batch:} $(s_n, a_n, r_n, s_{n+1})$ from $\mathcal{D}$
				\State \textbf{Critic Update:} 
				\State \quad Minimize value loss $L^{\text{Critic}}(\phi)$
				\State \textbf{Actor Update:}
				\State \quad Maximize hybrid PPO objective $L^{\text{Hybrid-PPO}}(\theta)$ 
				\Statex \quad \quad \quad \quad across all heads
				\EndFor
				\EndFor
			\end{algorithmic}
		\end{algorithm}
		
		\vspace{-2mm}
		\section{Simulation Results}
		\label{sim}
		In this section, we evaluate the performance of the proposed GATE-HPPO through extensive simulations. The simulation environment is carefully designed to reflect practical satellite-enabled service scenarios, considering both static and dynamic user distributions.
		\vspace{-2mm}
		\subsection{Simulation Environment}
		We consider a satellite-enabled seamless service system based on an MEO Rosette satellite constellation consisting of 8 satellites. The constellation is configured with eight orbital planes, each containing one satellite, deployed at an altitude of 20,184 km and an inclination of $53^{\circ}$. 
		The use of a MEO constellation is motivated by its larger coverage footprint and higher degree of regional overlap, providing a representative and challenging environment for evaluating migration-aware orchestration. In such settings, each satellite concurrently serves multiple user clusters with dynamic coverage intersections, facilitating clearer observation of service continuity and migration efficiency.
		Nevertheless, the proposed GATE-HPPO framework is architecture-agnostic and easily extendable to large-scale LEO constellations. Modeling the environment as a time-varying graph ensures that orbital altitude only influences the rate of topology evolution, not the underlying decision logic. This design inherently adapts to the short visibility windows and frequent handovers of satellite networks while alleviating centralized processing bottlenecks and ensuring scalability.
		The service period $T$ is set to 300 minutes, discretized into $N = 60$ time slots. The carrier frequency for communication is set to 14 GHz, with a minimum elevation angle threshold $\theta_{\min} = 15^{\circ}$ to determine visibility constraints between satellites and users.
		
		To reflect the spatiotemporal heterogeneity of wide-area service demands, we do not rely on homogeneous Poisson point processes to model the distribution of GUs. 
		Instead, we adopt a more realistic approach by using the geographic locations of 34 major global cities as representative user clusters. 
		Task arrival rates $\lambda_u$ at each location are assigned based on national population statistics obtained from recent census data, effectively capturing the non-uniform intensity of service demands across regions. 
		Formally, the arrival rate of user cluster $i$ is given by: $\lambda_i = \lambda_0 \cdot \frac{P_i}{\bar{P}},$ where $P_i$ denotes the population of city $i$, $\bar{P}$ is the average population across the selected cities, and $\lambda_0$ is a baseline reference rate. 
		This mapping ensures that urban clusters with higher population densities generate proportionally higher task arrival rates compared to suburban or rural clusters. 
		
		In addition to static GUs, the simulation incorporates 6 FUs representing aircraft operating within the satellite coverage area. 
		For FUs, real trajectory data from publicly available flight datasets are incorporated. 
		Specifically, we selected representative commercial flights and extracted their complete trajectories, including longitude, latitude, and altitude over the entire flight duration, which ensures that the mobility of FUs is modeled based on realistic air traffic patterns rather than simplified assumptions. 
		In terms of service demand, we distinguish between different flight phases. 
		Once an aircraft reaches a stable altitude, the subsequent period is regarded as the cruising phase. 
		During this phase only, FUs are assumed to generate service requests, with no requests initiated during takeoff or landing. 
		This design reflects practical in-flight conditions, where communication and computation services become relevant primarily after the aircraft has entered stable cruising flight. 
		By integrating real trajectory data and phase-aware service modeling, both the mobility dynamics and intermittent service demands of FUs are captured, enhancing the fidelity of the simulation environment.

		Simulations are conducted on a computer with an Intel Core i7-13700K CPU, 32 GB RAM, and an NVIDIA GeForce RTX 4070 Ti SUPER GPU. The software platform is PyCharm, and the language is Python. The models are implemented and trained using the PyTorch framework. 
		 The graph convolutional layers adopt a two-layer GCN with 32 and 16 hidden dimensions, respectively. The temporal encoder applies one 1D convolutional layer with kernel size $k = 3$. 
		The subsequent shared fully-connected network has layer sizes of [1024, 512]. Each of the per-user output heads for discrete satellite selection and continuous resource allocation (CPU and bandwidth) employs a 256-unit hidden layer before producing the final outputs, with the continuous heads utilizing Softplus activations to parameterize Beta distributions.
		For the hybrid PPO, the clipping parameter is set to $\epsilon = 0.2$, and the learning rate is set to $0.0001$.
		\vspace{-3mm}
		\subsection{Performance Analysis }
		\begin{figure*}[t]
			\begin{center}
				\subfigure[reward]{\includegraphics[width=0.8\textwidth,height=4.2cm, keepaspectratio]{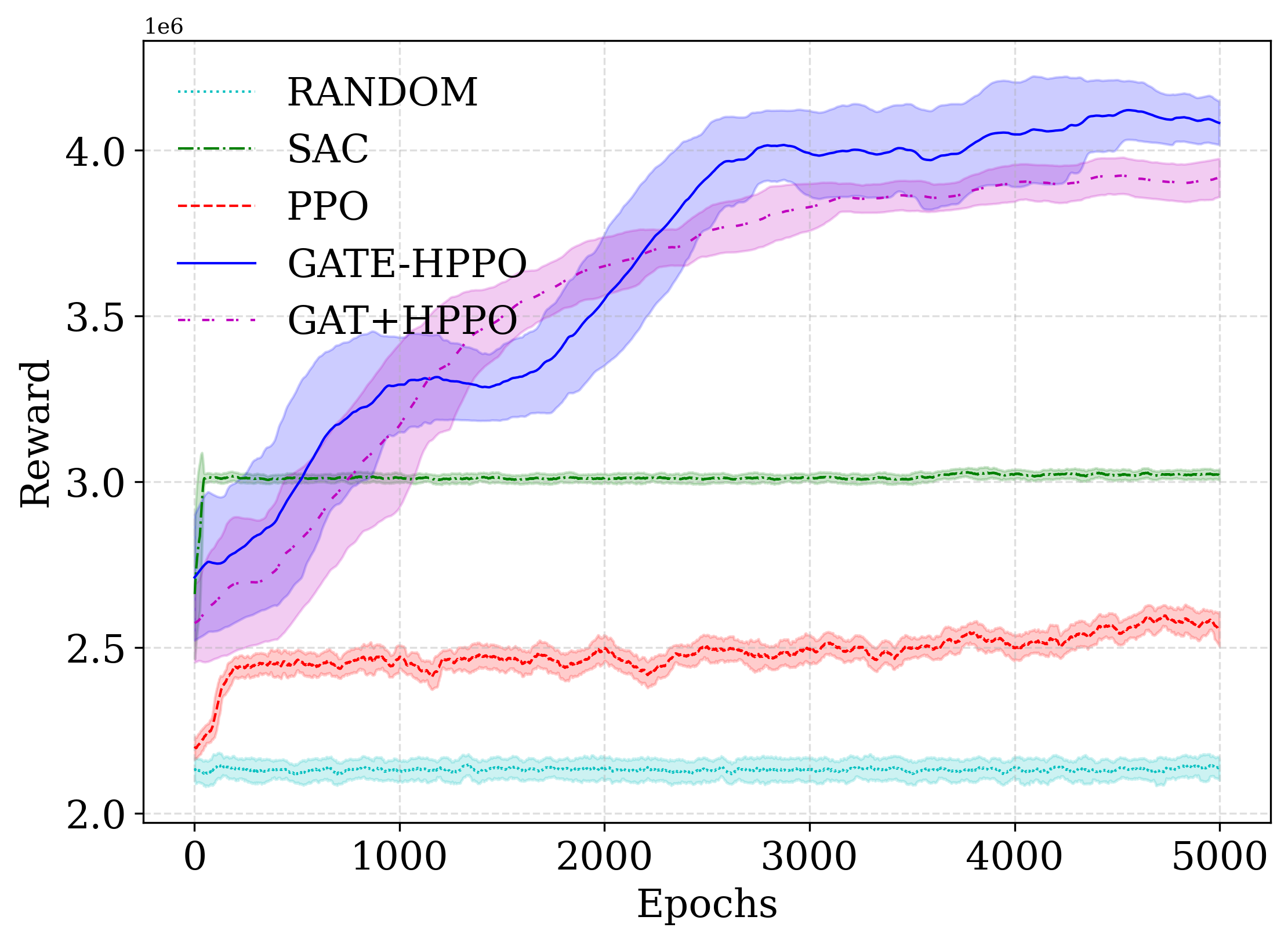}
					\label{fig:Performance_a}}\vspace{-1mm}
				\subfigure[Probability of unsuccessful service ]{\includegraphics[width=0.8\textwidth,height=4.2cm, keepaspectratio]{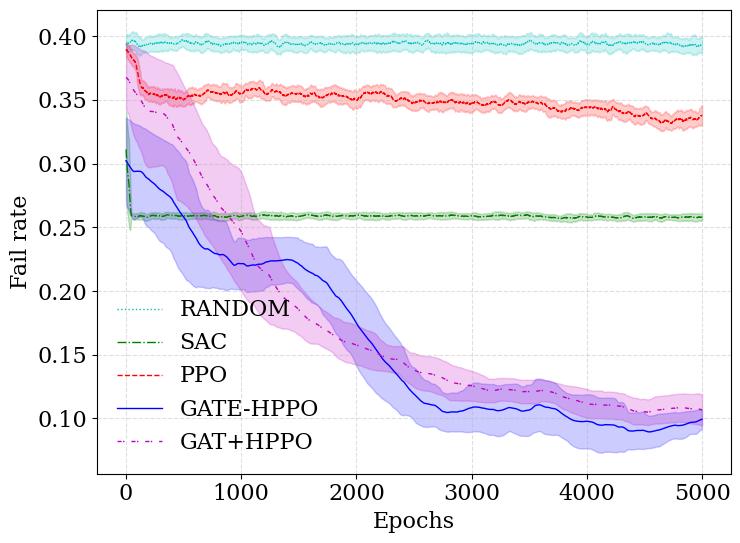}
					\label{fig:Performance_b}}\vspace{-1mm}
				\subfigure[Number of service migrations ]{\includegraphics[width=0.8\textwidth,height=4.2cm, keepaspectratio]{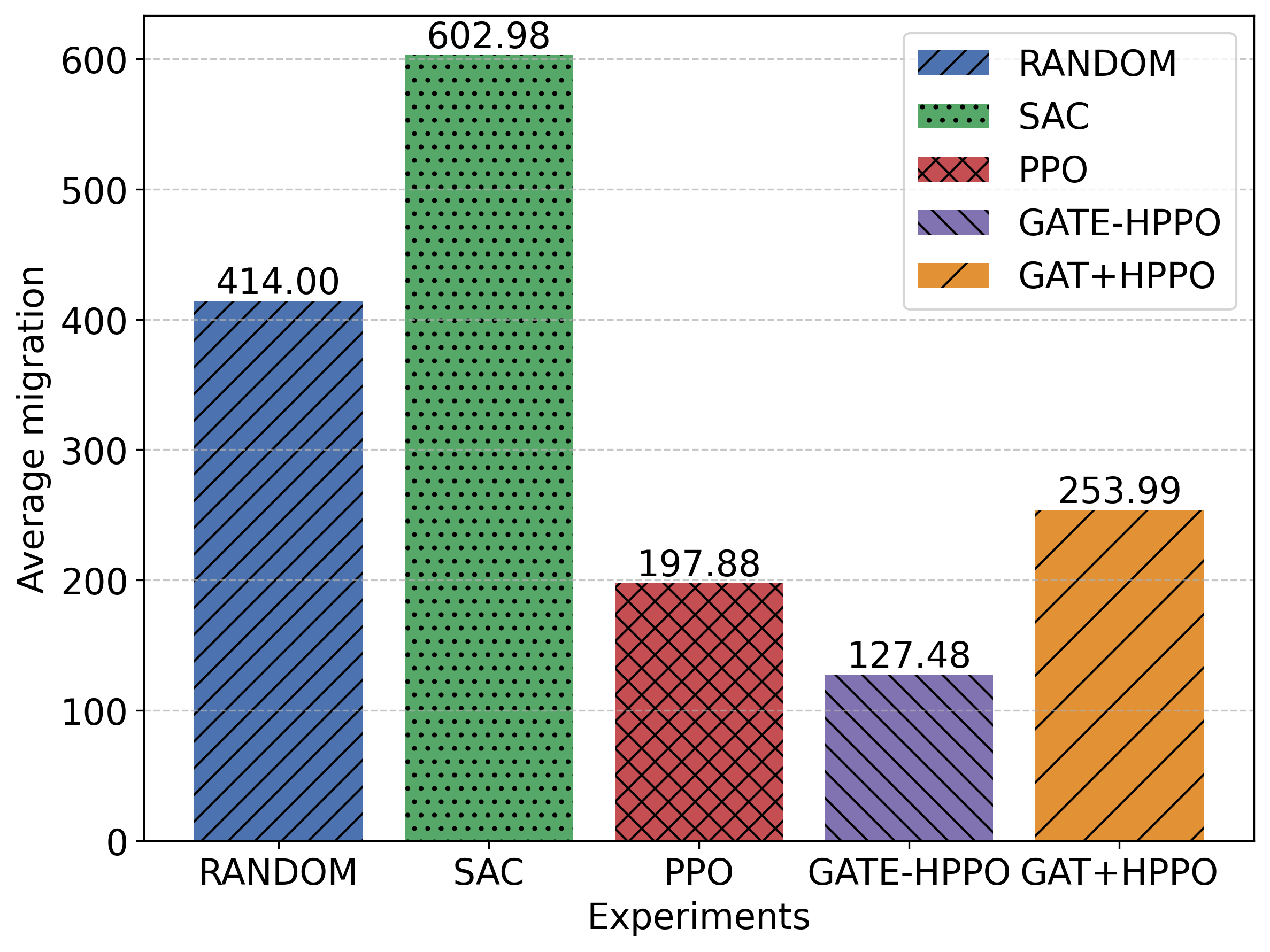}
					\label{fig:Performance_c}}\vspace{-1mm}			
				\caption{Performance of the proposed algorithm and the other baselines.}\vspace{-8mm}
				\label{fig:Performance}
			\end{center}
		\end{figure*}
		In this subsection, we comprehensively evaluate the performance of the proposed GATE-HPPO algorithm against three representative baselines: Random policy, PPO~\cite{schulman2017proximal}, Soft Actor-Critic (SAC)~\cite{haarnoja2018soft}, and a graph-enhanced DRL model (GAT+HPPO) that jointly captures spatial correlations and policy optimization. All algorithms are tested under identical environmental settings, with tuned hyperparameters to ensure fair comparison.
		The performance is assessed using three key metrics: reward, probability of unsuccessful service, and average number of service migrations, as illustrated in Figure~\ref{fig:Performance}.
		
		As shown in Fig.~\ref{fig:Performance_a}, GATE-HPPO achieves the highest accumulated reward, converging at approximately $4.12 \times 10^6$. In comparison, PPO, SAC, and GAT+HPPO reach $2.57\times10^6$, $3.02\times10^6$, and $3.89\times10^6$, respectively, corresponding to reward improvements of 60\%, 36\%, and 6.7\%. These results demonstrate that coupling graph convolutional feature extraction with temporal encoding enables more stable long-term performance than spatial modeling alone.
		Regarding service reliability, Fig.~\ref{fig:Performance_b} shows that GATE-HPPO achieves a failure rate below 10\% after convergence, substantially lower than PPO (35\%), SAC (25\%), and Random (38\%). Although GAT+HPPO attains a similar level (around 11\%), it requires a much higher number of migrations, indicating unstable policy behavior and excessive service switching. This underscores GATE-HPPO’s ability to balance reliability and operational efficiency through structured spatio-temporal state representation.
		As illustrated in Fig.~\ref{fig:Performance_c}, GATE-HPPO averages 127 service migrations, compared to 197 for PPO, 602 for SAC, 414 for Random, and 253 for GAT+HPPO, corresponding to reductions of 35.5\% and 49.8\% relative to PPO and GAT+HPPO, respectively. The significantly lower migration frequency demonstrates that GATE-HPPO’s temporal modeling, particularly through the sliding-window mechanism, enables proactive adaptation to mobility-induced topology changes while maintaining stable service continuity.
		
		Overall, GATE-HPPO consistently demonstrates superior performance across all evaluated metrics. These results validate the proposed framework's capability to leverage spatial-temporal structure in the environment and effectively optimize hybrid discrete-continuous decisions.
		\vspace{-1mm}
		\subsection{Per-User Reward and Service Demand Analysis}
		\begin{figure}[t]
			\centering
			\includegraphics[width=0.45\textwidth]{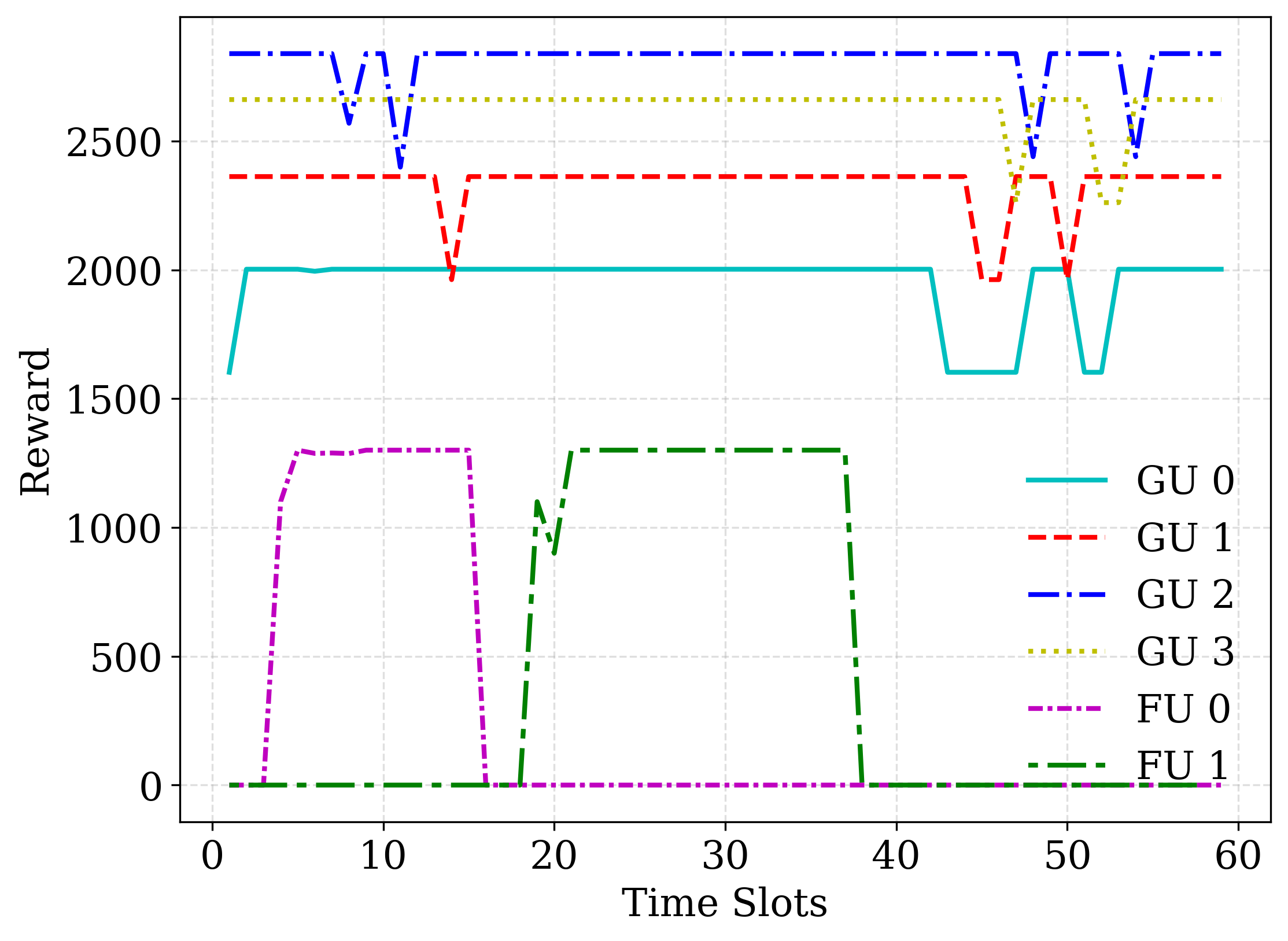}\vspace{-4mm}
			\caption{Per-user reward and service demand during the period.}\vspace{-4mm}
			\label{fig:user}
		\end{figure}
		To further investigate the algorithm’s adaptability to heterogeneous service demands, we analyze the reward dynamics of individual users over time, as presented in Figure~\ref{fig:user}. This analysis differentiates between GUs and FUs, highlighting variations in task arrival patterns and reward outcomes.
		
		GUs exhibit relatively stable reward profiles throughout the simulation horizon, consistent with their persistent and continuous service demands. Among the GUs, GU2 achieves the highest average reward, which aligns with its elevated task arrival rate and priority level. This observation confirms that the agent appropriately allocates resources in accordance with the joint influence of user importance and demand intensity.
		In contrast, FUs show intermittent reward patterns due to the time-constrained nature of their task generation, which occurs only during cruising phases. For instance, FU0 and FU1 produce non-zero rewards only during specific intervals, after takeoff and before landing. The highest reward for FU1 appears between time slots 20 and 40, aligning with its active service window. In general, the reward levels of FUs are lower than those of GUs, attributable to their shorter service durations and reduced request intensities.
		
		This per-user analysis demonstrates that GATE-HPPO dynamically adjusts its scheduling policy based on user heterogeneity. It ensures stable service performance for persistent GUs while flexibly accommodating dynamic and sporadic service demands from FUs. The results confirm the robustness and versatility of the proposed graph-aware temporal encoding framework in managing mixed-user satellite service systems.
		\vspace{-1mm}
		\subsection{Service Migration Dynamics Analysis}
\begin{figure}[t]
	\centering
	\includegraphics[width=0.45\textwidth]{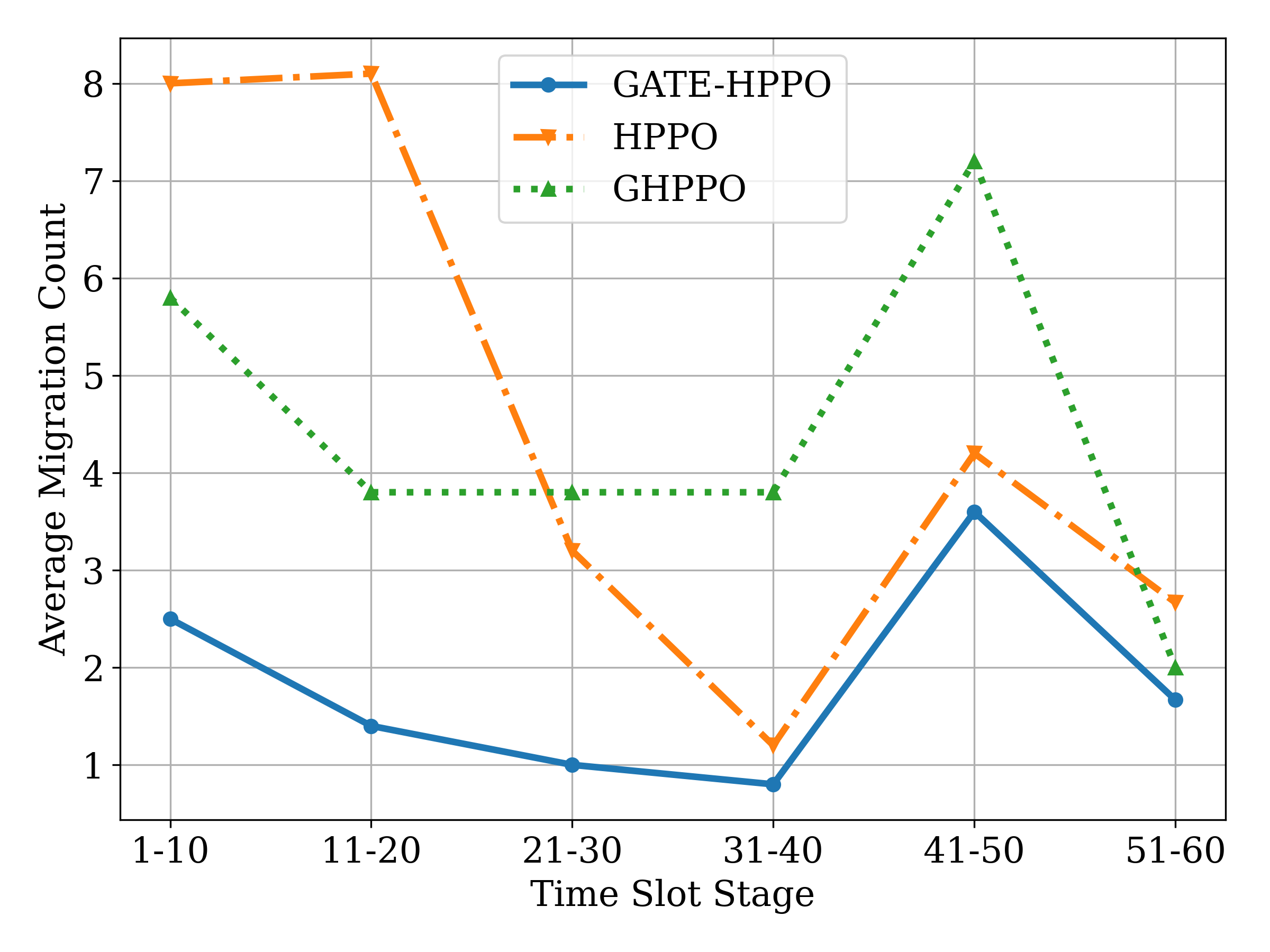}\vspace{-4mm}
	\caption{Migration Counts Over Time}\vspace{-6mm}
	\label{fig:migration}
\end{figure}

\begin{figure}[t]
			\centering
			\includegraphics[width=0.45\textwidth]{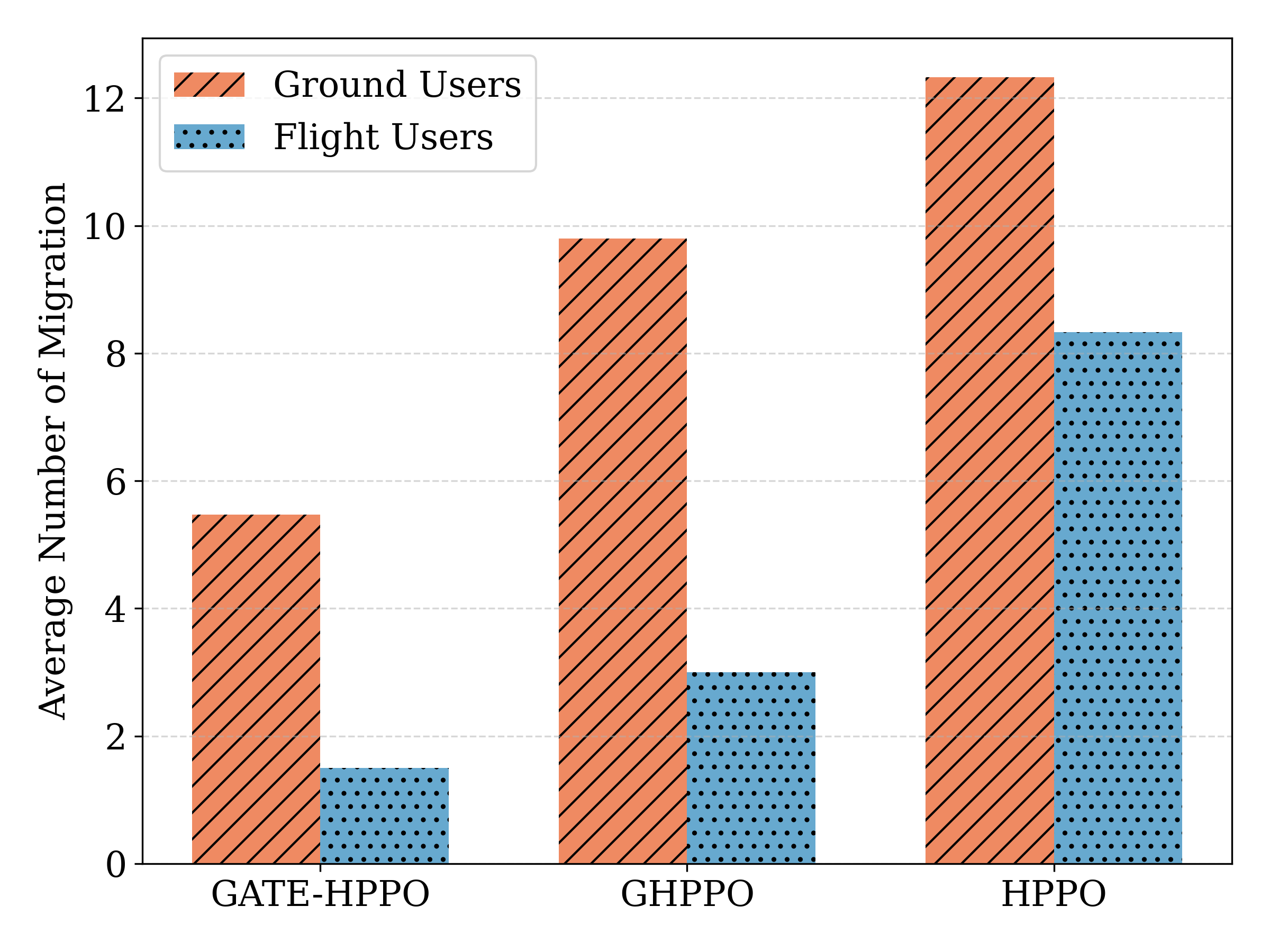}\vspace{-4mm}
			\caption{Average Migration Count}\vspace{-4mm}
			\label{fig:migration_user}
		\end{figure}

		To further examine the effectiveness of our proposed GATE-HPPO algorithm in managing service migration, we conduct a detailed analysis of migration behavior over time and across user categories. Fig.~\ref{fig:migration} and Fig.~\ref{fig:migration_user} provide insights into the temporal dynamics and user-specific characteristics of migration events under different methods.
		
		Figure~\ref{fig:migration} presents the average number of service migrations within six equal-length intervals across the entire scheduling period. Compared to HPPO and GHPPO, GATE-HPPO consistently exhibits both lower and more stable migration counts. HPPO displays a high concentration of migrations during the early stages, followed by a sharp decline. In contrast, GATE-HPPO maintains a gradual and steady migration trend throughout the scheduling horizon. This smoother pattern reflects a more deliberate and topology-aware reassignment strategy, effectively avoiding abrupt service relocations. GATE-HPPO helps mitigate network overhead and reduces latency variability, aligning better with the dynamic nature of satellite mobility and the requirement for continuous service delivery.
		
		Complementing this temporal view, Figure~\ref{fig:migration_user} presents a comparative analysis of the average number of service migrations experienced by GUs and FUs across different methods. The results demonstrate that GATE-HPPO consistently achieves lower migration counts for both user categories. Specifically, GUs experience an average of only 3.7 migrations under GATE-HPPO, which is markedly lower than the 8.9 and 11.4 observed under the two baseline methods. This indicates a substantial improvement in maintaining service continuity in areas with stable user demand. For FUs, whose mobility introduces greater scheduling complexity and frequent service disruptions, GATE-HPPO reduces the average number of migrations to 1.2, compared to 3.7 and 7.6 under the baselines. The pronounced reduction for both GUs and FUs highlights the algorithm's effectiveness in achieving stable and efficient service allocation, even in highly dynamic scenarios.
		
		Together, these findings underscore the robustness of GATE-HPPO in maintaining efficient and adaptive service reallocation strategies. By integrating spatial awareness with short-term temporal modeling, the algorithm effectively minimizes unnecessary migrations while preserving service continuity and optimizing network resource utilization.

		\vspace{-1mm}
		\subsection{Impact of Sliding Window Length}
		\begin{figure}[t]
			\centering
			\includegraphics[width=0.5\textwidth]{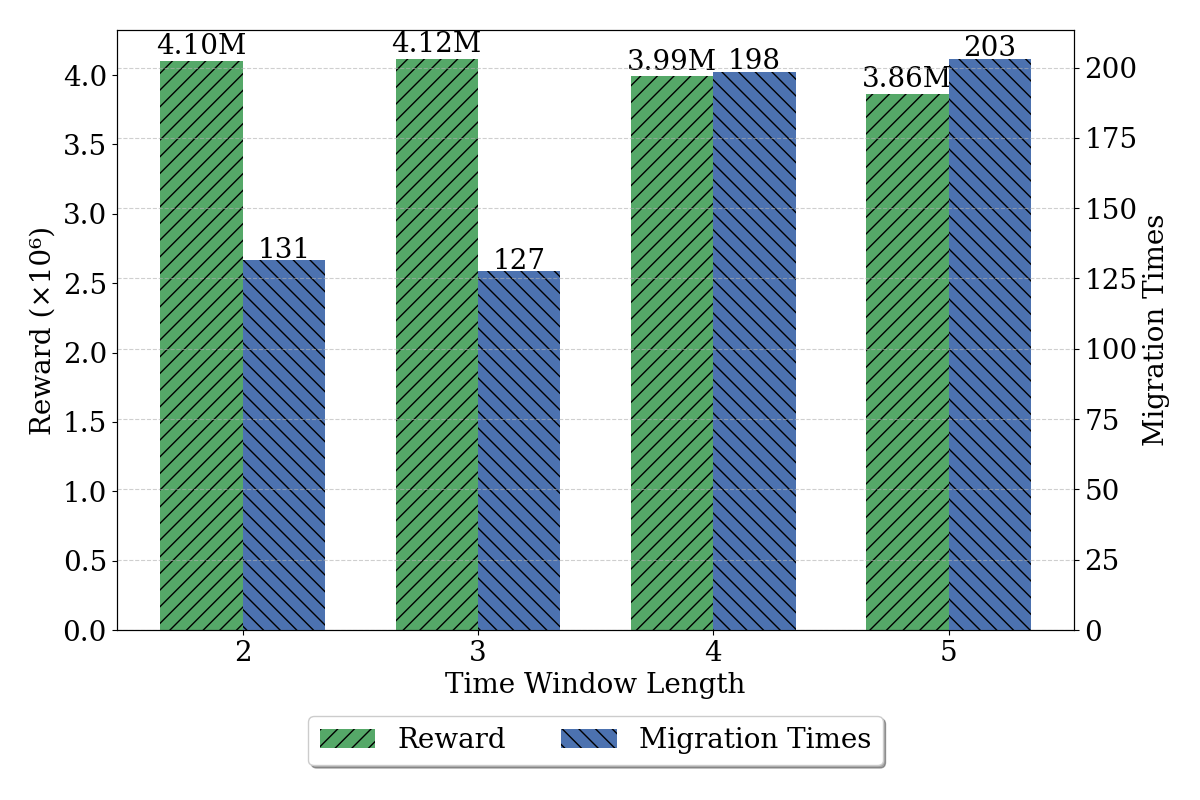}\vspace{-4mm}
			\caption{The effect of temporal encoding length on system performance.}\vspace{-6mm}
			\label{fig:time}
		\end{figure}
		
		To evaluate the effect of temporal encoding length on system performance, we conduct experiments varying the sliding window size $W$ from 2 to 5. As illustrated in Figure~\ref{fig:time}, the reward value and the number of service migrations both exhibit noticeable sensitivity to the choice of $W$. Specifically, when $W = 2$, the system achieves a cumulative reward of 4.1 million with an average migration count of 131. Increasing the window length to $W = 3$ further improves the reward to 4.12 million while reducing the migration count to 127, indicating that incorporating moderate temporal context enhances both system utility and migration stability.
		However, further enlarging the window length beyond $W = 3$ introduces diminishing returns and even adverse effects. At $W = 4$, the reward drops to 3.99 million, accompanied by a sharp increase in migration count to 198. When the window extends to $W = 5$, the degradation continues, with the reward decreasing to 3.86 million and migration count rising to 203. This suggests that overly long temporal windows may introduce outdated or redundant information, reducing the policy's responsiveness to the current system state.
		
		In summary, the results demonstrate that there exists an optimal temporal window length that balances historical context and decision agility. In our experiments, $W = 3$ yields the best trade-off between maximizing reward and minimizing unnecessary service migrations.

		\begin{figure*}[t]
			\begin{center}
				\subfigure[reward]{\includegraphics[width=0.8\textwidth,height=4.3cm, keepaspectratio]{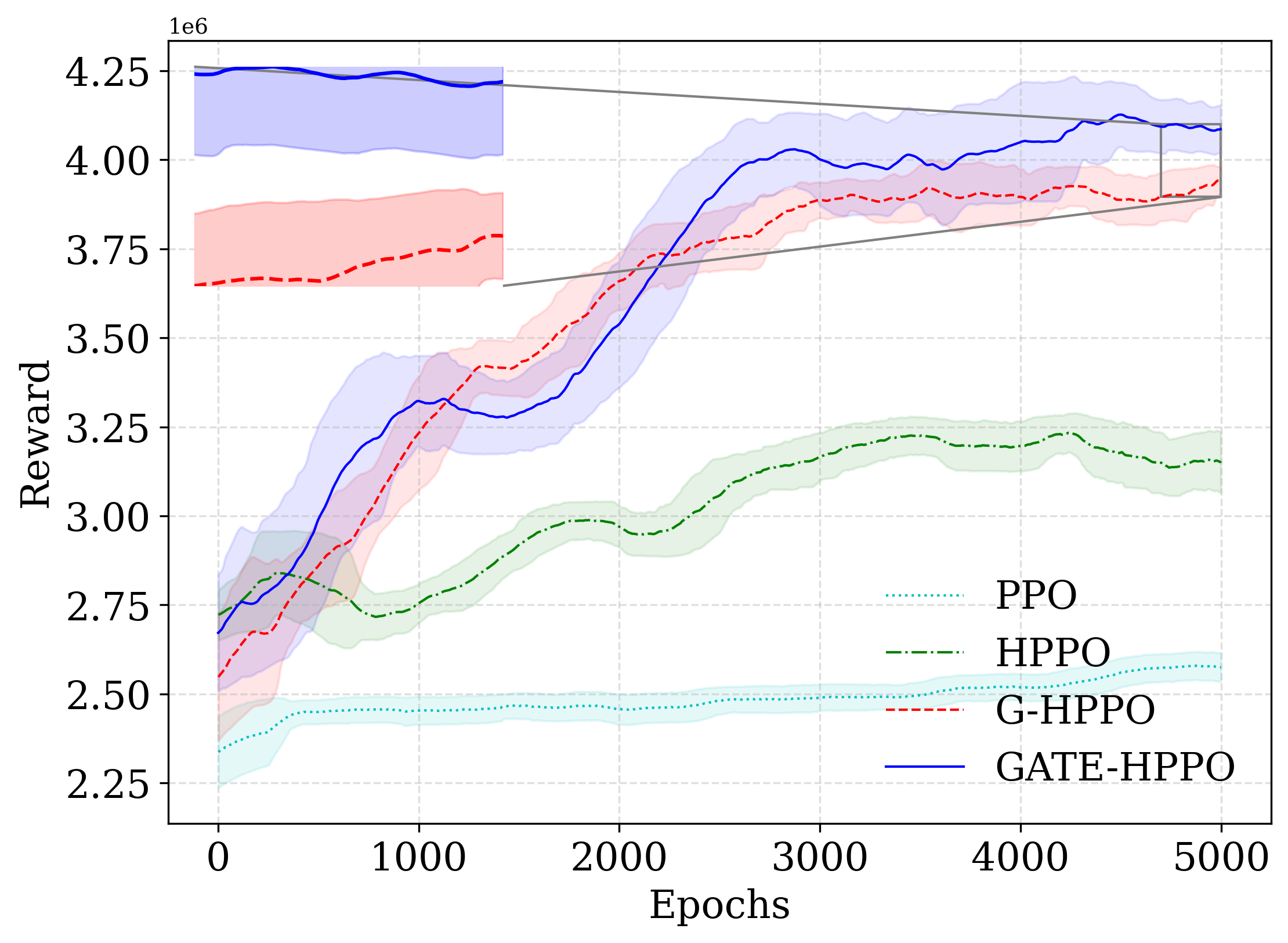}
					\label{fig:ablation_a}}\vspace{-1mm}
				\subfigure[Probability of unsuccessful service ]{\includegraphics[width=0.8\textwidth,height=4.3cm, keepaspectratio]{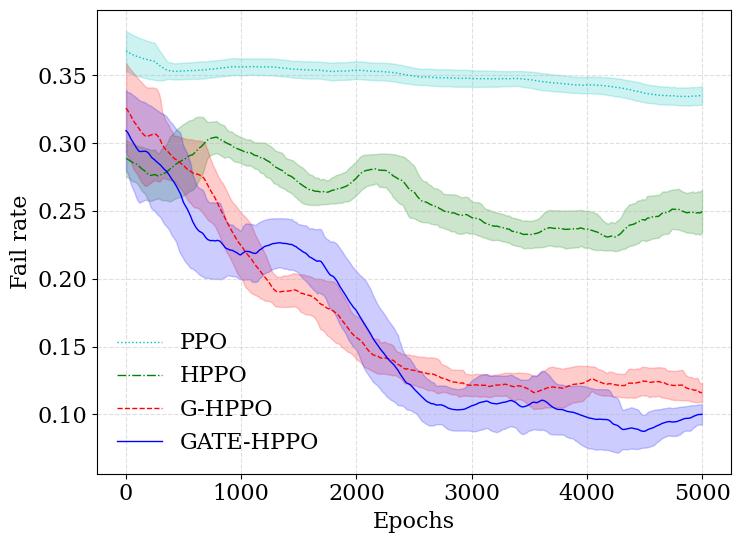}
					\label{fig:ablation_b}}\vspace{-1mm}
				\subfigure[Number of service migrations ]{\includegraphics[width=0.8\textwidth,height=4.3cm, keepaspectratio]{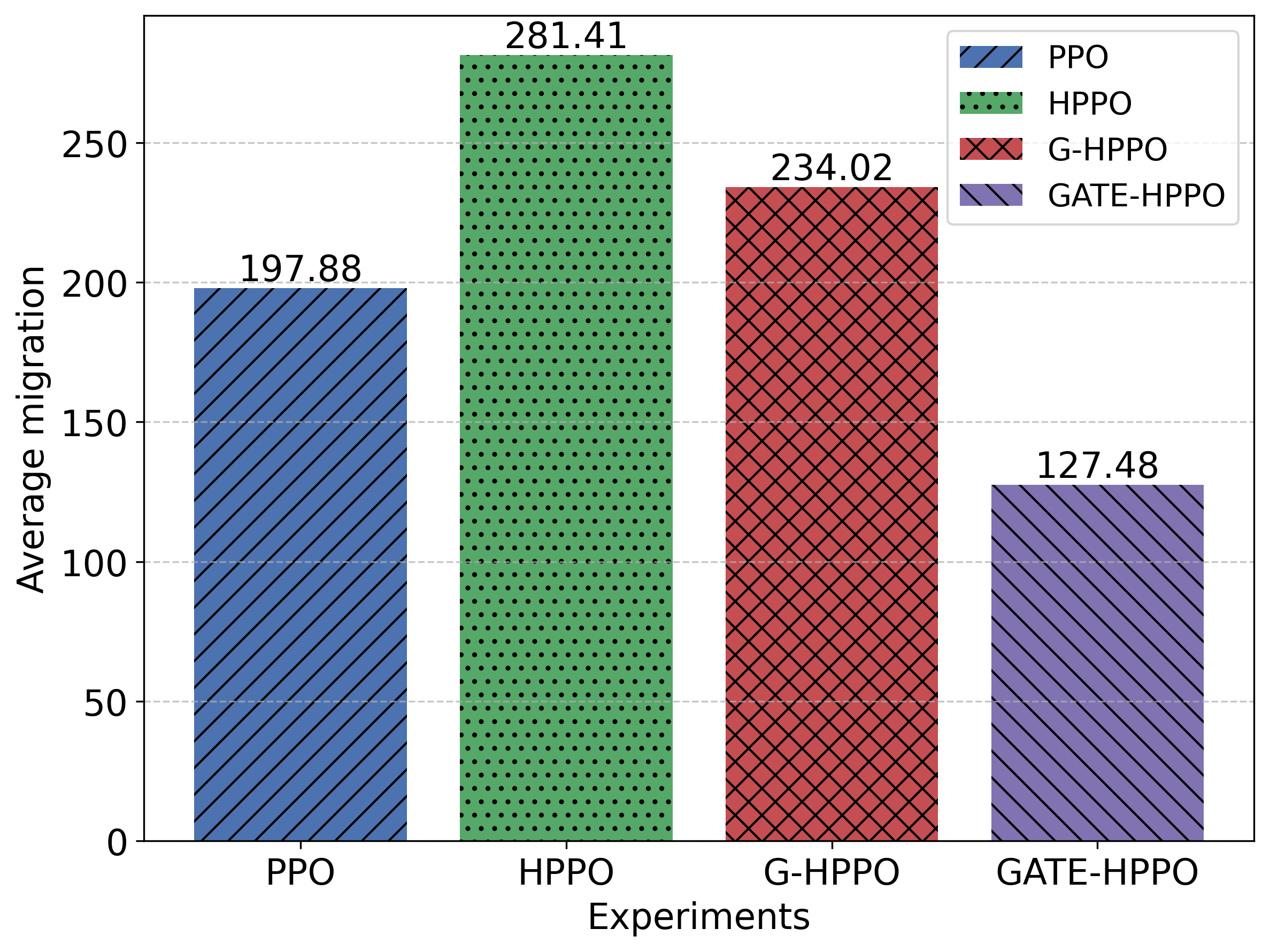}
					\label{fig:ablation_c}}\vspace{-1mm}			
				\caption{Ablation Experiment.}\vspace{-8mm}
				\label{fig:ablation}
			\end{center}
		\end{figure*}
		\begin{figure}[t]
			\centering
			\includegraphics[width=0.5\textwidth]{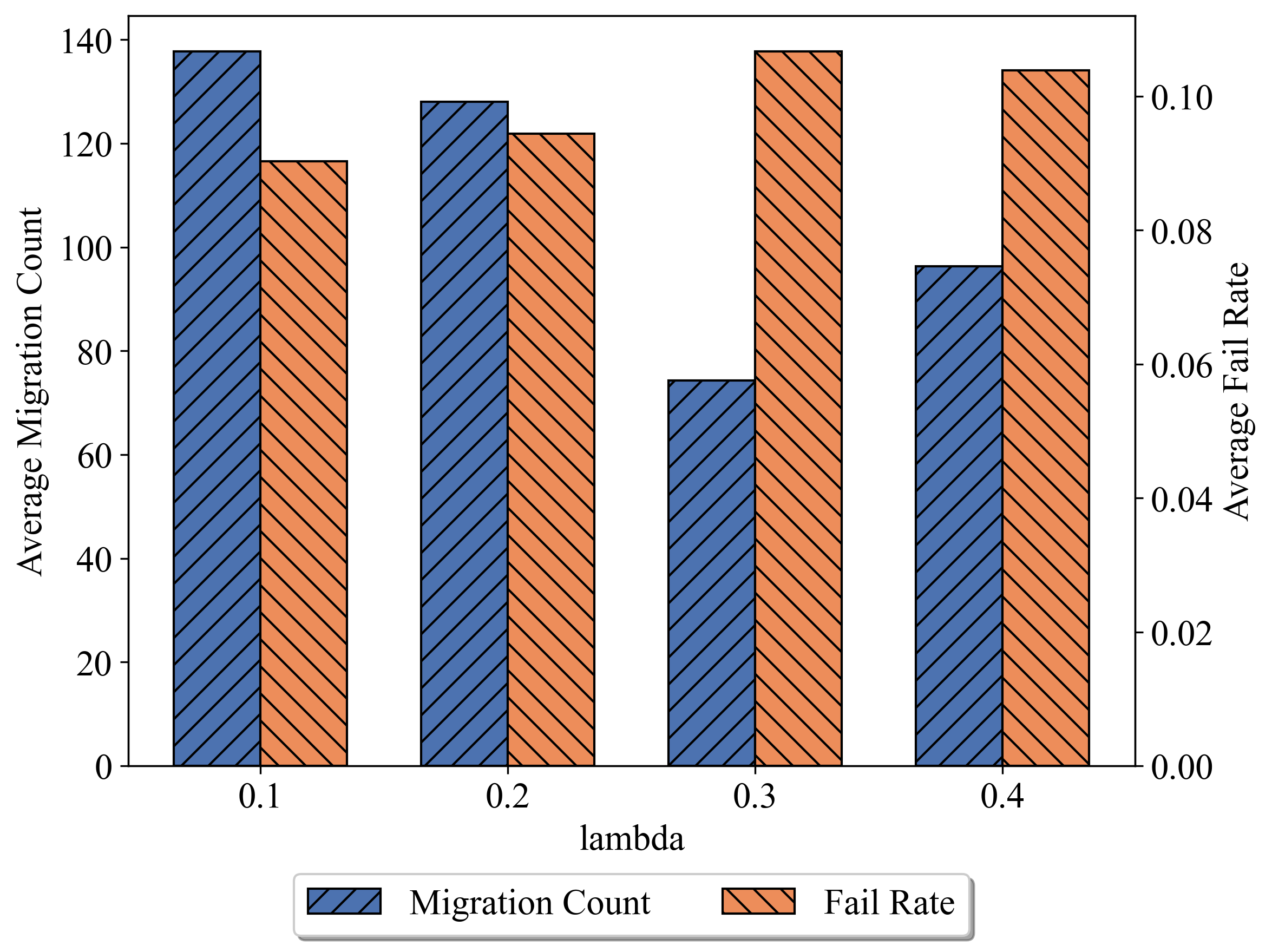}\vspace{-4mm}
			\caption{Parameter Sensitivity.}\vspace{-6mm}
			\label{fig:para}
			\vspace{-1mm}
		\end{figure}
		\vspace{-1mm}
		\subsection{Ablation Experiments}
		To systematically evaluate the contribution of each core component in our proposed GATE-HPPO framework, we conduct ablation experiments by progressively incorporating key modules and comparing their impact on system performance. Specifically, we consider the following variants: the standard PPO with discrete action space only, HPPO incorporating continuous resource allocation, Graph-based Hybrid PPO (GHPPO) adding graph convolutional feature extraction, and the full GATE-HPPO model with both graph convolution and temporal encoding.
		The first variant adopts the vanilla PPO algorithm, where the agent only selects a discrete satellite for service deployment without considering continuous resource allocation. The second variant, HPPO, extends PPO by incorporating a hybrid action space, enabling simultaneous optimization of satellite selection and continuous CPU and bandwidth allocation. The third variant, GHPPO, further enhances HPPO by introducing GCN to capture spatial dependencies between users and satellites. Finally, GATE-HPPO integrates temporal feature extraction through a sliding window mechanism based on temporal convolutional networks, allowing the policy to account for historical resource dynamics and user behaviors.
		
		Figure~\ref{fig:ablation} illustrates the comparative performance across three core metrics:  reward, probability of unsuccessful service, and number of service migrations. As shown in Figure~\ref{fig:ablation_a}, PPO achieves an average reward of approximately 2.57. Introducing continuous action heads in HPPO increases the reward to around 3.14, demonstrating the importance of flexible resource allocation. GHPPO further raises the reward to about 3.94, as graph-based spatial feature extraction enables more informed and accurate scheduling decisions. The complete GATE-HPPO model achieves the highest reward level at approximately 4.12. While the improvement from GHPPO to GATE-HPPO in reward appears small, this is primarily because the temporal encoding mainly affects stability rather than immediate utility.
		Figure~\ref{fig:ablation_b} compares the probability of unsuccessful service. GATE-HPPO consistently achieves the lowest failure rate, particularly under dynamic or bursty demand patterns such as those introduced by FUs. This validates the robustness of the proposed method in adapting to non-stationary conditions.
		Figure~\ref{fig:ablation_c} highlights differences in service migration. PPO records an average of 197 migrations over the last 100 training epochs. HPPO increases this to 281 due to its lack of temporal context. Adding spatial structure via GCNs reduces the count to 234 in GHPPO. Notably, GATE-HPPO lowers the average migration count to 127, demonstrating the effectiveness of temporal modeling in maintaining stable resource assignments and reducing operational overhead.
		
		Overall, the ablation experiments clearly demonstrate that each component of GATE-HPPO contributes uniquely to improving system performance. Hybrid action modeling enhances flexibility, graph convolutional feature extraction improves spatial awareness, and temporal encoding ensures stability and migration efficiency, jointly enabling the system to achieve superior efficiency, reliability, and adaptability.

		\vspace{-1mm}
		\subsection{Parameter Sensitivity}
		\vspace{-1mm}
		To further evaluate the robustness of GATE-HPPO, we analyze the impact of the reward weight $\lambda_{\text{pen}}$, which balances service quality and migration cost. As shown in Fig.~\ref{fig:para}, the average migration count initially decreases as $\lambda_{\text{pen}}$ increases from 0.1 to 0.3, since stronger penalization discourages unnecessary migrations. However, when $\lambda_{\text{pen}}$ exceeds 0.3, the migration count begins to rise again as over-penalization limits adaptive relocation and causes queue instability. Meanwhile, the service failure rate exhibits an increase with larger $\lambda_{\text{pen}}$, as overly conservative migration decisions reduce service continuity under satellite mobility. These results confirm that GATE-HPPO can flexibly adjust its behavior according to the penalty weight, and that $\lambda_{\text{pen}} = 0.2$ achieves the best trade-off between migration overhead and service reliability. Moreover, we observe that the overall performance remains stable under small perturbations of other RL parameters (e.g., learning rate and clipping threshold), demonstrating the robustness of our framework to hyperparameter variation.

		\vspace{-3mm}
		\section{Conclusion}
		\label{conculsion}
		
		In this paper, we presented a unified reinforcement learning framework that jointly optimizes service placement and resource allocation in dynamic satellite networks. By modeling the system as a time-varying graph and incorporating realistic queuing effects, our formulation captures both spatial connectivity variations and heterogeneous user demand patterns. The proposed GATE distills rich spatio-temporal features, while the HPPO architecture efficiently handles the mixed discrete–continuous action space. Extensive simulations with both ground and flight users show that our approach achieves a 36\% increase in accumulated reward, a 60\% reduction in service failure rate, and 35\% fewer service migrations compared with baseline methods. These quantitative results underscore the effectiveness of leveraging structured graph representations and short-term temporal context in orchestrating edge services for satellite networks.
%

		%
		\vspace{-2mm}
		\appendices
		\bibliographystyle{IEEEtran}
		\bibliography{ref}

\begin{thebibliography}{10}
\providecommand{\url}[1]{#1}
\csname url@samestyle\endcsname
\providecommand{\newblock}{\relax}
\providecommand{\bibinfo}[2]{#2}
\providecommand{\BIBentrySTDinterwordspacing}{\spaceskip=0pt\relax}
\providecommand{\BIBentryALTinterwordstretchfactor}{4}
\providecommand{\BIBentryALTinterwordspacing}{\spaceskip=\fontdimen2\font plus
\BIBentryALTinterwordstretchfactor\fontdimen3\font minus
  \fontdimen4\font\relax}
\providecommand{\BIBforeignlanguage}[2]{{%
\expandafter\ifx\csname l@#1\endcsname\relax
\typeout{** WARNING: IEEEtran.bst: No hyphenation pattern has been}%
\typeout{** loaded for the language `#1'. Using the pattern for}%
\typeout{** the default language instead.}%
\else
\language=\csname l@#1\endcsname
\fi
#2}}
\providecommand{\BIBdecl}{\relax}
\BIBdecl

\bibitem{fontanesi2025artificial}
G.~Fontanesi, F.~Ort{\'\i}z, E.~Lagunas, L.~M. Garc{\'e}s-Socarr{\'a}s, V.~M.
  Baeza, M.~{\'A}. V{\'a}zquez, J.~A. V{\'a}squez-Peralvo, M.~Minardi, H.~N.
  Vu, P.~J. Honnaiah \emph{et~al.}, ``Artificial intelligence for satellite
  communication: A survey,'' \emph{IEEE Commun. Surv. Tutor. (Early Access)},
  2025.

\bibitem{tang2024joint}
Q.~Tang, R.~Xie, Z.~Fang, T.~Huang, T.~Chen, R.~Zhang, and F.~R. Yu, ``Joint
  service deployment and task scheduling for satellite edge computing: A
  two-timescale hierarchical approach,'' \emph{IEEE J. Sel. Areas Commun.},
  vol.~42, no.~5, pp. 1063--1079, Feb. 2024.

\bibitem{9444334}
Q.~Li, S.~Wang, X.~Ma, Q.~Sun, H.~Wang, S.~Cao, and F.~Yang, ``Service coverage
  for satellite edge computing,'' \emph{I{EEE} Internet Things J.}, vol.~9,
  no.~1, pp. 695--705, Jan. 2022.

\bibitem{wang2024resource}
G.~Wang, F.~Yang, J.~Song, and Z.~Han, ``Resource allocation and load balancing
  for beam hopping scheduling in satellite-terrestrial communications: A
  cooperative satellite approach,'' \emph{IEEE Trans. Wirel. Commun.}, vol.~24,
  no.~2, pp. 1339--1354, Feb. 2024.

\bibitem{gao2024hierarchical}
X.~Gao, Y.~Hu, Y.~Shao, H.~Zhang, Y.~Liu, R.~Liu, and J.~Zhang, ``Hierarchical
  dynamic resource allocation for computation offloading in {LEO} satellite
  networks,'' \emph{IEEE Internet Things J.}, vol.~11, no.~11, pp.
  19\,470--19\,484, Jun. 2024.

\bibitem{mao2024intelligent}
B.~Mao, X.~Zhou, J.~Liu, and N.~Kato, ``On an intelligent hierarchical routing
  strategy for ultra-dense free space optical low earth orbit satellite
  networks,'' \emph{IEEE J. Sel. Areas Commun.}, vol.~42, no.~5, pp.
  1219--1230, Feb. 2024.

\bibitem{zhao2025demand}
R.~Zhao, J.~Cai, J.~Luo, J.~Gao, and Y.~Ran, ``Demand-aware beam hopping and
  power allocation for load balancing in digital twin empowered {LEO} satellite
  networks,'' \emph{IEEE Trans. Wirel. Commun.}, vol.~24, no.~6, pp.
  5084--5098, Jun. 2025.

\bibitem{maity2024traffic}
I.~Maity, G.~Giambene, T.~X. Vu, C.~Kesha, and S.~Chatzinotas, ``Traffic-aware
  resource management in {SDN}/{NFV}-based satellite networks for remote and
  urban areas,'' \emph{IEEE Trans. Veh. Technol.}, vol.~73, no.~11, pp.
  17\,400--17\,415, Nov. 2024.

\bibitem{zakarya2024apmove}
M.~Zakarya, L.~Gillam, A.~A. Khan, O.~Rana, and R.~Buyya, ``Apmove: A service
  migration technique for connected and autonomous vehicles,'' \emph{IEEE
  Internet Things J.}, vol.~11, no.~17, pp. 28\,721--28\,733, Sep. 2024.

\bibitem{ji2024dynamic}
S.~Ji, D.~Zhou, M.~Sheng, J.~Li, and Z.~Han, ``Dynamic space-ground integrated
  mobility management strategy for mega {LEO} satellite constellations,''
  \emph{IEEE Trans. Wirel. Commun.}, vol.~23, no.~9, pp. 11\,043--11\,060, Sep.
  2024.

\bibitem{du2022sdnTON}
J.~Du, C.~Jiang, A.~Benslimane, S.~Guo, and Y.~Ren, ``S{DN}-based resource
  allocation in edge and cloud computing systems: An evolutionary stackelberg
  differential game approach,'' \emph{IEEE/ACM Trans. Networking}, vol.~30,
  no.~4, pp. 1613--1628, Aug. 2022.

\bibitem{jiang2025federated}
W.~Jiang, J.~Mu, H.~Han, Y.~Zhang, and S.~Huang, ``Federated learning-based
  mobile traffic prediction in satellite-terrestrial integrated networks,''
  \emph{Softw. Pract. Exp.}, vol.~55, no.~4, pp. 613--628, Nov. 2025.

\bibitem{razmi2024board}
N.~Razmi, B.~Matthiesen, A.~Dekorsy, and P.~Popovski, ``On-board federated
  learning for satellite clusters with inter-satellite links,'' \emph{IEEE
  Trans. Commun.}, vol.~72, no.~6, pp. 3408--3424, Jun. 2024.

\bibitem{lin2024fedsn}
Z.~Lin, Z.~Chen, Z.~Fang, X.~Chen, X.~Wang, and Y.~Gao, ``Fedsn: A federated
  learning framework over heterogeneous {LEO} satellite networks,'' \emph{IEEE
  IEEE. Trans. Mob. Comput.}, vol.~24, no.~3, pp. 1293--1307, Mar. 2024.

\bibitem{yang2024dfedsat}
\BIBentryALTinterwordspacing
M.~Yang, J.~Zhang, and S.~Liu, ``Dfedsat: Communication-efficient and robust
  decentralized federated learning for {LEO} satellite constellations,'' 2024.
  [Online]. Available: \url{https://arxiv.org/abs/2407.05850}
\BIBentrySTDinterwordspacing

\bibitem{shi2024satellite}
Y.~Shi, L.~Zeng, J.~Zhu, Y.~Zhou, C.~Jiang, and K.~B. Letaief, ``Satellite
  federated edge learning: Architecture design and convergence analysis,''
  \emph{IEEE Trans. Wirel. Commun.}, vol.~23, no.~10, pp. 15\,212--15\,229,
  Oct. 2024.

\bibitem{gong2024multi}
Y.~Gong, H.~Yao, Z.~Xiong, D.~Yu, X.~Cheng, C.~Yuen, M.~Bennis, and M.~Debbah,
  ``Multi-modal federated learning based resources convergence for
  satellite-ground twin networks,'' \emph{IEEE. Trans. Mob. Comput.}, vol.~24,
  no.~5, pp. 4104--4117, Dec. 2024.

\bibitem{zhang2024collaborative}
H.~Zhang, H.~Zhao, R.~Liu, A.~Kaushik, X.~Gao, and S.~Xu, ``Collaborative task
  offloading optimization for satellite mobile edge computing using multi-agent
  deep reinforcement learning,'' \emph{IEEE Trans. Veh. Technol.}, vol.~7,
  no.~10, pp. 15\,483--15\,498, Oct. 2024.

\bibitem{zhao2024qos}
L.~Zhao, Y.~Liu, A.~Hawbani, N.~Lin, W.~Zhao, and K.~Yu, ``Qos-aware multihop
  task offloading in satellite--terrestrial edge networks,'' \emph{IEEE
  Internet Things J.}, vol.~11, no.~19, pp. 31\,453--31\,466, Oct. 2024.

\bibitem{zhou2024latency}
J.~Zhou, J.~Liang, L.~Zhao, S.~Wan, H.~Cai, and F.~Xiao, ``Latency-energy
  efficient task offloading in the satellite network-assisted edge computing
  via deep reinforcement learning,'' \emph{IEEE. Trans. Mob. Comput.}, vol.~24,
  no.~4, pp. 2644--2659, Apr. 2024.

\bibitem{zhong2025joint}
L.~Zhong, Y.~Li, M.-F. Ge, M.~Feng, and S.~Mao, ``Joint task offloading and
  resource allocation for {LEO} satellite-based mobile edge computing systems
  with heterogeneous task demands,'' \emph{IEEE Trans. Veh. Technol.}, vol.~74,
  no.~7, pp. 11\,337--11\,352, Jul. 2025.

\bibitem{gong2025multi}
Y.~Gong, D.~Yu, H.~Yao, X.~Cheng, A.~Nallanathan, and G.~K. Karagiannidis,
  ``Multi-modal learning based multi-task offloading schemes for
  satellite-ground integrated networks,'' \emph{IEEE Trans. Wirel. Commun.},
  vol.~24, no.~7, pp. 5635--5647, Jul. 2025.

\bibitem{cai2024dynamic}
Y.~Cai, P.~Cheng, Z.~Chen, W.~Xiang, B.~Vucetic, and Y.~Li, ``Dynamic resource
  management with graphic deep reinforcement learning in space-air-ground
  integrated networks,'' in \emph{Proc. IEEE Glob. Commun. Conf.
  (GLOBECOM)}.\hskip 1em plus 0.5em minus 0.4em\relax Cape Town, South Africa:
  IEEE, 8-12 Dec. 2024, pp. 1491--1496.

\bibitem{du2025collaborative}
X.~Du, Z.~Na, N.~Zhang, Y.~Zhang, M.~Jia, and Z.~Gao, ``Collaborative task
  offloading for leo satellite internet of things: A novel computing coordinate
  graph-based approach,'' \emph{IEEE Sys. J.}, 2025, early access, doi:
  {10.1109/JSYST.2025.3571386}.

\bibitem{duj2025CM}
J.~Du, H.~Wang, C.~Jiang, J.~Simonjan, J.~Wang, and M.~Debbah, ``Distributed
  {AI}-based secure communications in space-air-ground-sea integrated
  networks,'' \emph{IEEE Commun. Mag.}, vol.~63, no.~7, pp. 48--55, Jul. 2025.

\bibitem{wang2024dynamic}
H.~Wang, Y.~Gao, Z.~Guo, L.~Yan, and S.~Cao, ``Dynamic service migration
  mechanism in satellite edge computing with location privacy protection,'' in
  \emph{Proc. IEEE Int. Conf. Commun. Technol. (ICCT)}, Chengdu, China, Oct.,
  2024.

\bibitem{chen2024spaceedge}
J.-H. Chen, W.-C. Kuo, and W.~Liao, ``Spaceedge: Optimizing service latency and
  sustainability for space-centric task offloading in {LEO} satellite
  networks,'' \emph{IEEE Trans. Wirel. Commun.}, vol.~23, no.~10, pp.
  15\,435--15\,446, Oct. 2024.

\bibitem{li2023online}
Z.~Li, H.~Zhang, C.~Liu, X.~Li, H.~Ji, and V.~C. Leung, ``Online service
  deployment on mega-{LEO} satellite constellations for end-to-end delay
  optimization,'' \emph{IEEE Trans. Netw. Sci. Eng.}, vol.~11, no.~1, pp.
  1214--1226, Oct. 2023.

\bibitem{li2021distributed}
Z.~Li, C.~Jiang, and J.~Lu, ``Distributed service migration in satellite mobile
  edge computing,'' in \emph{Proc. IEEE Global Commun. Conf. (GLOBECOM)},
  Madrid, Spain, Dec., 2021.

\bibitem{deng2023bandwidth}
P.~Deng, X.~Gong, and X.~Que, ``A bandwidth-aware service migration method in
  {LEO} satellite edge computing network,'' \emph{Comput. Commun.}, vol. 200,
  pp. 104--112, Feb. 2023.

\bibitem{cao2022edge}
X.~Cao, B.~Yang, Y.~Shen, C.~Yuen, Y.~Zhang, Z.~Han, H.~V. Poor, and L.~Hanzo,
  ``Edge-assisted multi-layer offloading optimization of {LEO}
  satellite-terrestrial integrated networks,'' \emph{IEEE J. Sel. Areas
  Commun.}, vol.~41, no.~2, pp. 381--398, Dec. 2022.

\bibitem{liu2024multi}
H.~Liu, Y.~Wang, P.~Li, and J.~Cheng, ``A multi-agent deep reinforcement
  learning-based handover scheme for mega-constellation under dynamic
  propagation conditions,'' \emph{IEEE Wirel. Commun.}, vol.~23, no.~10, pp.
  13\,579--13\,596, Oct. 2024.

\bibitem{yang2024multi}
F.~Yang, W.~Wu, Y.~Gao, Y.~Sun, T.~Sun, and P.~Si, ``Multi-agent
  fingerprints-enhanced distributed intelligent handover algorithm in {LEO}
  satellite networks,'' \emph{IEEE Trans. Veh. Technol.}, vol.~73, no.~10, pp.
  15\,255--15\,269, Oct. 2024.

\bibitem{recommendation2005838}
``Specific attenuation model for rain for use in prediction methods,''
  \emph{International Telecommunication Union: Geneva, Switzerland},
  ITU-Recommendation 838–3 2005.

\bibitem{recommendation2023618}
``Propagation data and prediction methods required for the design of
  earth-space telecommunication systems,'' \emph{International
  Telecommunication Union: Geneva, Switzerland}, ITU-Recommendation P.618-14
  2023.

\bibitem{zhou2018channel}
D.~Zhou, M.~Sheng, R.~Liu, Y.~Wang, and J.~Li, ``Channel-aware mission
  scheduling in broadband data relay satellite networks,'' \emph{IEEE J. Sel.
  Areas Commun.}, vol.~36, no.~5, pp. 1052--1064, May 2018.

\bibitem{gongora2022link}
J.~M. Gongora-Torres, C.~Vargas-Rosales, A.~Arag{\'o}n-Zavala, and
  R.~Villalpando-Hernandez, ``Link budget analysis for {LEO} satellites based
  on the statistics of the elevation angle,'' \emph{IEEE Access}, vol.~10, pp.
  14\,518--14\,528, Jan. 2022.

\bibitem{vu2021dynamic}
T.~X. Vu, S.~Chatzinotas, and B.~Ottersten, ``Dynamic bandwidth allocation and
  precoding design for highly-loaded multiuser miso in beyond 5g networks,''
  \emph{IEEE Trans. Wirel. Commun.}, vol.~21, no.~3, pp. 1794--1805, Mar. 2021.

\bibitem{wang2022bandwidth}
J.~Wang, X.~Zhang, X.~He, and Y.~Sun, ``Bandwidth allocation and trajectory
  control in uav-assisted iov edge computing using multiagent reinforcement
  learning,'' \emph{IEEE Trans. Reliab.}, vol.~72, no.~2, pp. 599--608, Jun.
  2022.

\bibitem{wang2018survey}
S.~Wang, J.~Xu, N.~Zhang, and Y.~Liu, ``A survey on service migration in mobile
  edge computing,'' \emph{IEEE Access}, vol.~6, pp. 23\,511--23\,528, Apr.
  2018.

\bibitem{machen2017live}
A.~Machen, S.~Wang, K.~K. Leung, B.~J. Ko, and T.~Salonidis, ``Live service
  migration in mobile edge clouds,'' \emph{IEEE Wirel. Commun.}, vol.~25,
  no.~1, pp. 140--147, Feb. 2017.

\bibitem{huang2024aoi}
J.~Huang, T.~Yu, F.~Yang, S.~Zhang, W.~Jiang, and D.~Niyato, ``Ao{I}-aware
  resource allocation with interference avoidance for ultra-dense industrial
  {I}nternet of {T}hings networks,'' \emph{IEEE Internet Things J.}, vol.~11,
  no.~17, pp. 28\,787--28\,797, May 2024.

\bibitem{du2024fmWCM}
J.~Du, T.~Lin, C.~Jiang, Q.~Yang, C.~F. Bader, and Z.~Han, ``Distributed
  foundation models for multi-modal learning in 6{G} wireless networks,''
  \emph{IEEE Wireless Commun.}, vol.~31, no.~3, pp. 20--30, Jun. 2024.

\bibitem{cai2024graphic}
Y.~Cai, P.~Cheng, Z.~Chen, W.~Xiang, B.~Vucetic, and Y.~Li, ``Graphic deep
  reinforcement learning for dynamic resource allocation in space-air-ground
  integrated networks,'' \emph{IEEE J. Sel. Areas Commun.}, vol.~43, no.~1, pp.
  334--349, Jan. 2024.

\bibitem{dong2021intelligent}
T.~Dong, Z.~Zhuang, Q.~Qi, J.~Wang, H.~Sun, F.~R. Yu, T.~Sun, C.~Zhou, and
  J.~Liao, ``Intelligent joint network slicing and routing via gcn-powered
  multi-task deep reinforcement learning,'' \emph{IEEE Trans. Cognit. Commun.
  Networking}, vol.~8, no.~2, pp. 1269--1286, Jun. 2021.

\bibitem{schulman2017proximal}
\BIBentryALTinterwordspacing
J.~Schulman, F.~Wolski, P.~Dhariwal, A.~Radford, and O.~Klimov, ``Proximal
  policy optimization algorithms,'' 2017. [Online]. Available:
  \url{http://arxiv.org/abs/1707.06347}
\BIBentrySTDinterwordspacing

\bibitem{haarnoja2018soft}
T.~Haarnoja, A.~Zhou, P.~Abbeel, and S.~Levine, ``Soft actor-critic: Off-policy
  maximum entropy deep reinforcement learning with a stochastic actor,'' in
  \emph{Proc. Int. Conf. Mach. Learn. (ICML)}, Stockholm , Sweden, Jul., 2018.

\end{thebibliography}
		\vspace{-2mm}
		\begin{IEEEbiography}[{\includegraphics[width=1in,height=1.25in,clip,keepaspectratio]{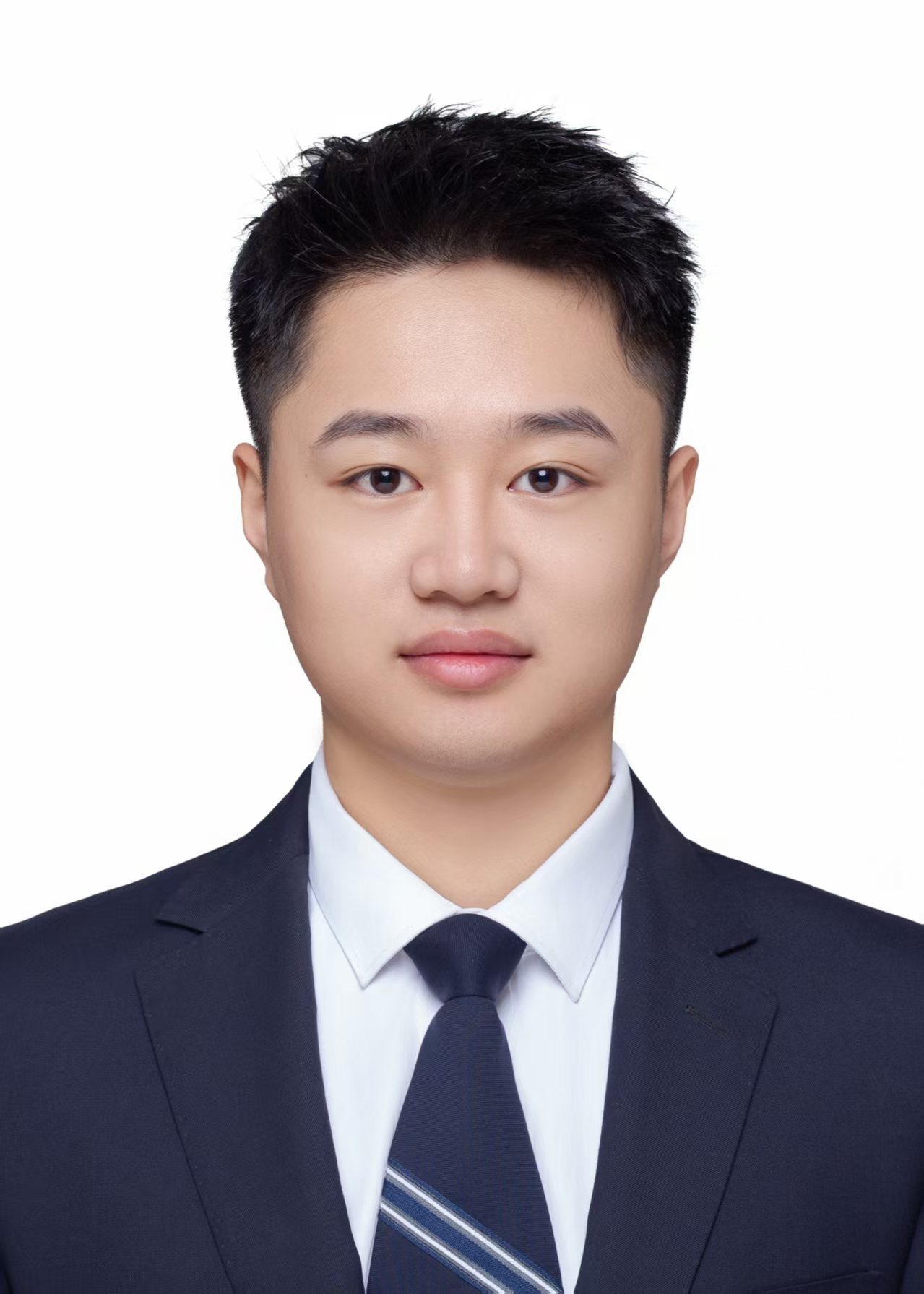}}]{Haotong Wang} (Graduate Student Member, IEEE) received his B.S. degree from the School of Microelectronics and Communication Engineering, Chongqing University, Chongqing, China, in 2024. He is currently pursuing the M.S. degree in electronics and communication engineering with Tsinghua University, Beijing. His research interests include machine learning and resource allocation.
		\end{IEEEbiography}
		\vspace{-2mm}
		\begin{IEEEbiography}[{\includegraphics[width=1in,height=1.25in,clip,keepaspectratio]{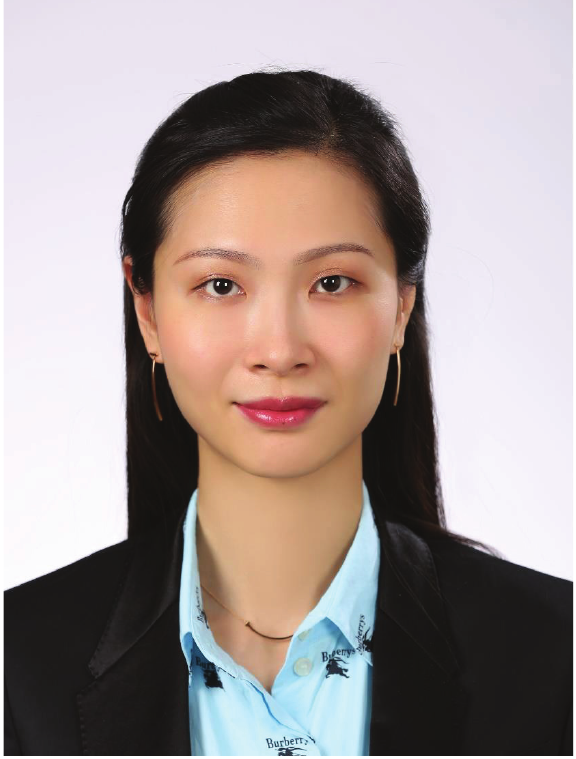}}]{Jun Du} (Senior Member, IEEE) received her B.S. in information and communication engineering from Beijing Institute of Technology, in 2009, and her M.S. and Ph.D. in information and communication engineering from Tsinghua University, Beijing, in 2014 and 2018, respectively. From Oct. 2016- Sept. 2017, Dr. Du was a sponsored researcher, and she visited Imperial College London. Currently she is an associate professor in the Department of Electrical Engineering, Tsinghua University. Her research interests are mainly in communications, networking, resource allocation and system security problems of heterogeneous networks and space-based information networks. Dr. Du is the recipient of the Best Student Paper Award from IEEE GlobalSIP in 2015, the Best Paper Award from IEEE ICC 2019 and 2025, and the Best Paper Award from IWCMC in 2020.
		\end{IEEEbiography}
		
		\vspace{-2mm}
		\begin{IEEEbiography}[{\includegraphics[width=1in,height=1.25in,clip,keepaspectratio]{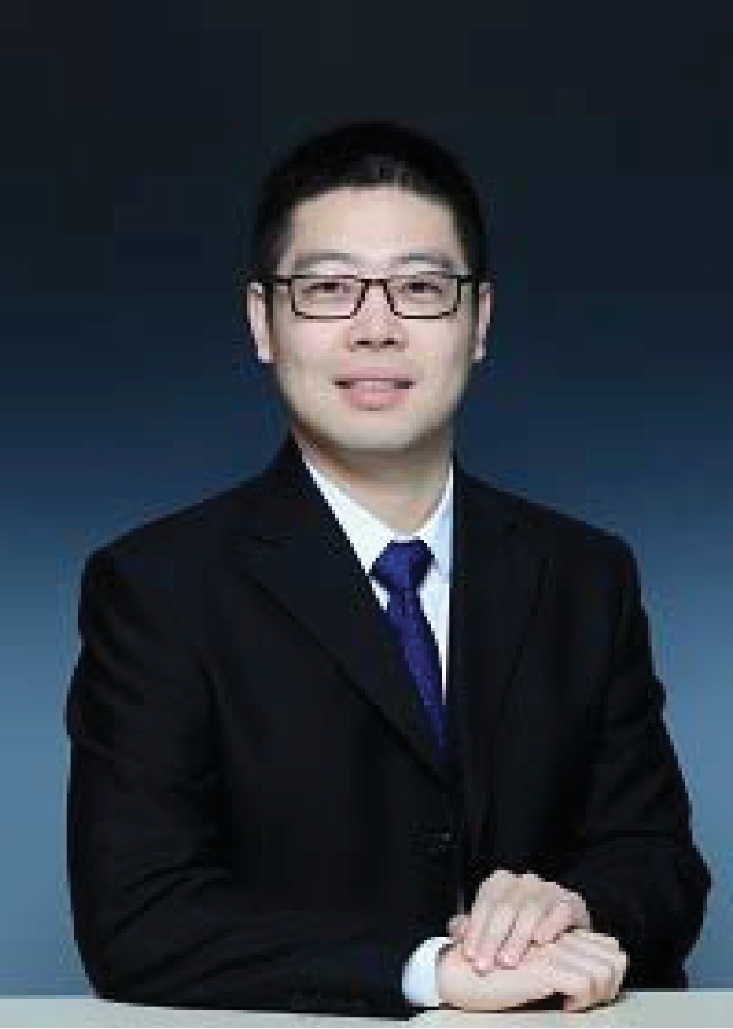}}]{Chunxiao Jiang} (Fellow, IEEE) is an associate professor in School of Information Science and Technology, Tsinghua University. He received the B.S. degree in information engineering from Beihang University, Beijing in 2008 and the Ph.D. degree in electronic engineering from Tsinghua University, Beijing in 2013, both with the highest honors. From 2011 to 2012 (as a Joint Ph.D) and 2013 to 2016 (as a Postdoc), he was in the Department of Electrical and Computer Engineering at University of Maryland College Park under the supervision of Prof. K. J. Ray Liu. His research interests include application of game theory, optimization, and statistical theories to communication, networking, and resource allocation problems, in particular space networks and heterogeneous networks. Dr. Jiang is the recipient of the Best Paper Award from IEEE GLOBECOM in 2013, IEEE Communications Society Young Author Best Paper Award in 2017, the Best Paper Award from ICC 2019, IEEE VTS Early Career Award 2020 IEEE ComSoc Asia-Pacific Best Young Researcher Award 2020, IEEE VTS Distinguished Lecturer 2021, and IEEE ComSoc Best Young Professional Award in Academia 2021. He received the Chinese National Second Prize in Technical Inventions Award in 2018. He is a Fellow of IEEE and a Fellow of IET.
		\end{IEEEbiography}
		\vspace{-2mm}
		\begin{IEEEbiography}[{\includegraphics[width=1in,height=1.25in,clip,keepaspectratio]{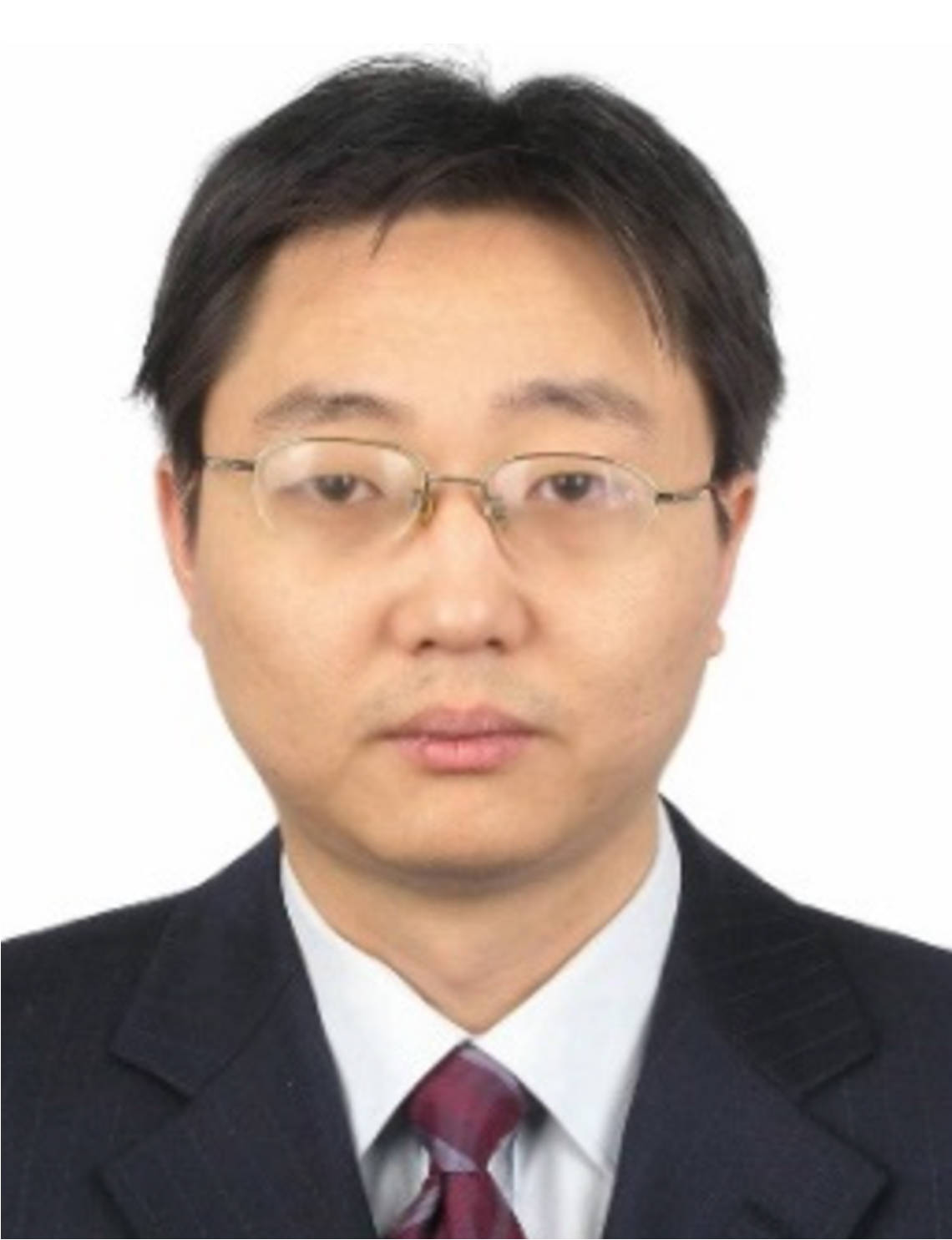}}]{Jintao Wang} (Senior Member, IEEE) received the B.Eng. and Ph.D. degrees in electrical engineering from Tsinghua University, Beijing, China, in 2001 and 2006, respectively. From 2006 to 2009, he was an Assistant Professor with the Department of Electronic Engineering, Tsinghua University. Since 2009, he has been an Associate Professor and a Ph.D. Supervisor, and he was promoted to a Full Professor, in 2019. He is the Standard Committee Member for the Chinese national digital terrestrial television broadcasting standard. He has published more than 100 journal and conference papers and holds more than 50 national invention patents. His current research interests include space-time coding, MIMO, and OFDM systems.
		\end{IEEEbiography}
		\vspace{-2mm}
		\begin{IEEEbiography}[{\includegraphics[width=1in,height=1.25in,clip,keepaspectratio]{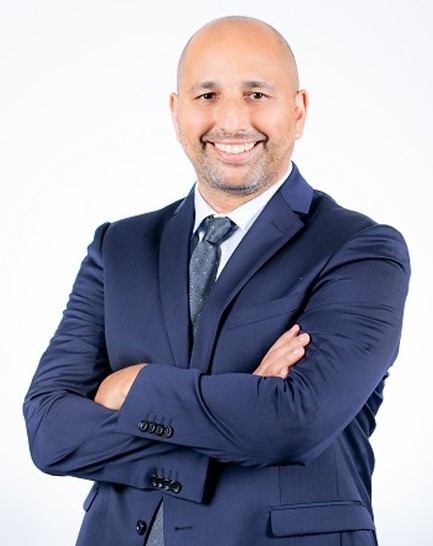}}]{M\'{e}rouane Debbah} (Fellow, IEEE) is currently a Professor at the Khalifa University of Science and Technology, Abu Dhabi, and the Founding Director of the KU 6G Research Center. He is a frequent keynote speaker at international events in the field of telecommunication and AI. His research has been lying at the interface of fundamental mathematics, algorithms, statistics, information, and communication sciences with a special focus on random matrix theory and learning algorithms. In the communication field, he has been at the heart of the development of small cells (4G), massive MIMO (5G), and large intelligent surfaces (6G) technologies. In the AI field, he is known for his work on large language models, distributed AI systems for networks, and semantic communications. He received multiple prestigious distinctions, prizes, and best paper awards (more than 50 IEEE best paper awards) for his contributions to both fields. He is a fellow of WWRF, Eurasip, AAIA, Institut Louis Bachelier, and AIIA; and a Membre \'{E}m\'{e}rite SEE. He is the Chair of the IEEE Large Generative AI Models in Telecom (GenAINet) Emerging Technology Initiative and a member of the Marconi Prize Selection Advisory Committee.
		\end{IEEEbiography}
		\vspace{-2mm}
		\begin{IEEEbiography}[{\includegraphics[width=1in,height=1.25in,clip,keepaspectratio]{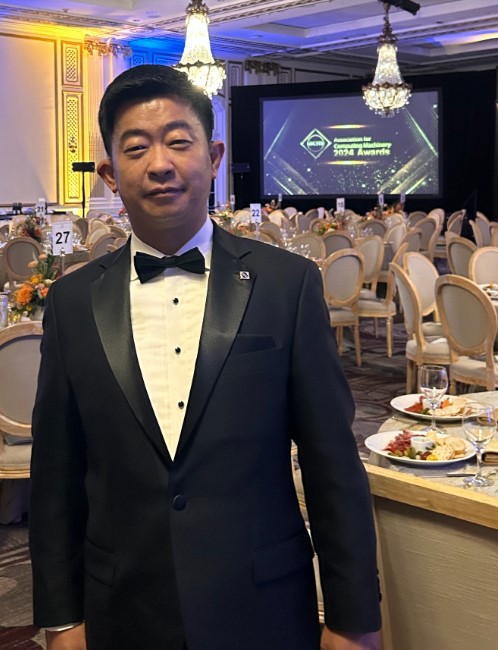}}]{Zhu Han} (Fellow, IEEE) received the B.S. degree in electronic engineering from Tsinghua University, in 1997, and the M.S. and Ph.D. degrees in electrical and computer engineering from the University of Maryland, College Park, in 1999 and 2003, respectively. 
		From 2000 to 2002, he was an R\&D Engineer of JDSU, Germantown, Maryland. From 2003 to 2006, he was a Research Associate at the University of Maryland. From 2006 to 2008, he was an assistant professor at Boise State University, Idaho. Currently, he is a John and Rebecca Moores Professor in the Electrical and Computer Engineering Department as well as in the Computer Science Department at the University of Houston, Texas. Dr. Han’s main research targets on the novel game-theory related concepts critical to enabling efficient and distributive use of wireless networks with limited resources. His other research interests include wireless resource allocation and management, wireless communications and networking, quantum computing, data science, smart grid, carbon neutralization, security and privacy.  Dr. Han received an NSF Career Award in 2010, the Fred W. Ellersick Prize of the IEEE Communication Society in 2011, the EURASIP Best Paper Award for the Journal on Advances in Signal Processing in 2015, IEEE Leonard G. Abraham Prize in the field of Communications Systems (best paper award in IEEE JSAC) in 2016, IEEE Vehicular Technology Society 2022 Best Land Transportation Paper Award, and several best paper awards in IEEE conferences. Dr. Han was an IEEE Communications Society Distinguished Lecturer from 2015 to 2018 and ACM Distinguished Speaker from 2022 to 2025, AAAS fellow since 2019, and ACM Fellow since 2024. Dr. Han is also the winner of the 2021 IEEE Kiyo Tomiyasu Award (an IEEE Field Award), for outstanding early to mid-career contributions to technologies holding the promise of innovative applications, with the following citation: ``for contributions to game theory and distributed management of autonomous communication networks."		
		\end{IEEEbiography}

	\end{document}